\documentclass[openany]{revtex4-2}
\usepackage{graphicx}
\usepackage{xcolor}

\newcommand{\rf}[1]{(\ref{#1})}

\newcommand{\beq}{\begin{equation}}
\newcommand{\eeq}{\end{equation}}
\newcommand{\beqr}{\begin{eqnarray}}
\newcommand{\eeqr}{\end{eqnarray}}
\newcommand{\lb}[1]{\label{#1}}
\newcommand{\bc}{\begin{center}}
\newcommand{\ec}{\end{center}}
\newcommand{\ct}[1]{\cite{#1}}

\begin{document}
\title{Quantum fluctuations in the  small Fabry-Perot interferometer }
\author{Igor E. Protsenko}
\email{procenkoie@lebedev.ru}

\author{Alexander V. Uskov}

\affiliation{%
Quantum Electronic division, Lebedev Physical Institute,
Moscow 119991, Russia
}

\date{\today}

\begin{abstract}We consider the small, of the size of the order of the wavelength, interferometer with the main mode  excited by a quantum field from a nano-LED or a laser. The input field is detuned from the
interferometer mode with, on average, a few photons. We find the field and the photon number fluctuation spectra inside and outside the interferometer and identify the contributions of quantum and classical noise in the spectra.  Structures of spectra are different for the field, the
photon number fluctuations inside the interferometer; for the transmitted, and the reflected fields. We note asymmetries in spectra. Differences
in the spectra are related to the colored (white) quantum noise inside (outside) the interferometer. We
calculate the second-order time correlation functions; they oscillate and be negative under certain
conditions. Results help the study, design, manufacture, and use small elements of quantum
optical integrated circuits, such as delay lines and optical transistors. 
\begin{description}
\item[Keywords] quantum noise, interferometer, optical element, field spectrum
\end{description}
\end{abstract}

\maketitle

\section{Introduction}
Fabry-Perot interferometer (FPI), invented in 1899 \ct{FP_1899}, is widely used in optics, optoelectronics, and laser physics \ct{Vaughan1989,Ismail:16}. FPIs can be met, in particular,  in telecommunications for wavelength-division multiplexing \ct{KEISER19993}, as laser cavities \ct{trove.nla.gov.au/work/21304573}, in spectroscopy to control and measure the wavelengths of light \ct{2003102}, in precision displacement measurements (chapter 5.10.1.1 of \ct{Leach2014}), in optical integrated circuits  \ct{1159437,Zhu:21,Elshaari2020}. Recent technological progress leads to a considerable reduction of the optical element size \ct{Wang:16,Liu:18} and the appearance of photonic quantum technologies (PQT) \ct{Pelucchi2022}. PQT requires experimental research and theoretical studies of quantum phenomena in a small optical elements, with the size of the order of the optical wavelength, such as the small FPI, operating with a few photons  at a large amount of quantum noise. The paper  contributes to the quantum theory of such a small FPI.  

The particular motivation for the quantum consideration of the small FPI is in making the background for the theoretical  model of the small quantum, single-photon optical transistor for PQT.  It is well-known that the FPI with the nonlinear medium has, at certain conditions,   dispersive optical bistability \ct{doi:10.1080/00107518308210690, PhysRevA.19.2074,doi:10.1063/1.88632}, and operates as an optical transistor \ct{Bowden_book,PROTSENKO1994304}. So the bistable miniature FPI is an important element for the quantum photonic circuits, necessary for ultra-low power signal processing  \ct{Kerckhoff:11,4054411}. 

In this paper, we calculate, in particular, the photon number fluctuation spectrum of the FPI with the quantum input field detuned from the center of the FPI mode. It is a necessary step for analyzing the bistability in the quantum nonlinear FPI with only a few photons. We will do such analysis in the future with the method of \ct{Protsenko_2022}. This method permits solving  nonlinear operator equations and generalizes a cumulant-neglect closure approach of the classical stochastic theory \ct{Wu_1984,10.1115/1.3173083} to spectral analysis of open quantum nonlinear systems such as lasers and nonlinear optical devices. The cumulant-neglect closure approach has been used previously for quantum systems in the cluster expansion method  \ct{Jahnke,PhysRevA.75.013803} for calculations of high-order correlations.

Another motivation of the present study is the investigation of the field, the field power fluctuation spectra, and the auto-correlation functions of the small FPI with a mode detuned from the quantum input field. We find spectral profiles different from Lorentzian (or, Airy) spectral distributions well-known from the classical theory of FPI \ct{Ismail:16}. We see that spectra are asymmetric, with a noticeable contribution of quantum fluctuations, and different shapes for transmitted, reflected fields and the field inside the FPI. Previous quantum theory of  the FPI has considered, for example, the quantum limits of measurements \ct{doi:10.1080/09500348714550251} without particular attention to the FPI spectra. The incoherent input field has been applied to FPI in experiments, for example, for the characterization of optical Fabry-Perot cavities \ct{Tsuchida:12}.

We model the FPI as the quantum harmonic oscillator excited by the quantum stochastic force. The oscillator with a stochastic excitation is one of the basic models in the stochastic theory \ct{Gitterman2005,Oraevsky_1987}  and for the quantum case  \ct{PhysRevE.85.031110}. We hope that this paper contributes to the spectral  theory of open quantum harmonic oscillators and helps to extend the method of \ct{Protsenko_2022} from the laser theory to general quantum devices  \ct{Sutherland2003} modeled as sets of oscillators.

We present general formulas for the photon number fluctuation spectra inside the
FPI and the field power fluctuation spectra outside the FPI; formulate the model for the
FPI interacting with a quantum field and finding explicit expressions for the FPI spectra in
section~\ref{sec2}.

We show and describe spectra and auto-correlation functions of the FPI, interacting with the quantum field, in section \ref{sec3}.  

The discussion of the results of section \ref{sec3} is in section \ref{sec4}. 
\section{\label{sec2}Methods: formulas for spectra and quantum model of the FPI} 
\begin{figure}[thb]
\includegraphics[width=10.5 cm]{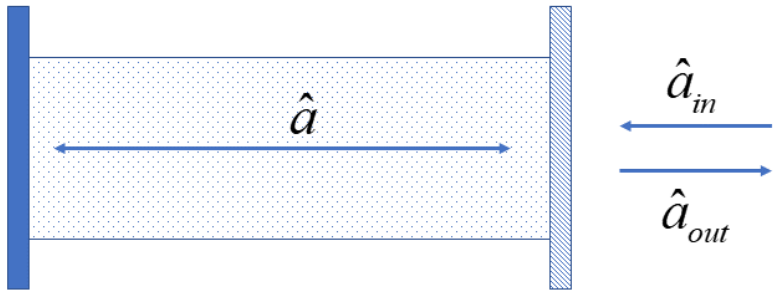}
\caption{Fabry-Perot interferometer with two semitransparent mirrors, the input (reflected) field amplitude operators $\hat{a}_{in}$ ($\hat{a}_r$) on the right, transmitted field $\hat{a}_t$ on the left, and the field $\hat{a}$ inside.}\label{fig1}
\end{figure}

Here we derive formulas for the photon number fluctuation (the field power) spectra for fields inside (outside) the FPI, whith the scheme shown in Fig.~\ref{fig1}.
%
%

%
%
\noindent Quazi-monochromatic external field with Bose-operator ${{\hat{a}}_{in}}(t){{e}^{-i{{\omega }_{l}}t}}$, with the amplitude operator ${{\hat{a}}_{in}}(t)$ and the carrier frequency $\omega_l$, enters the FPI, shown in  Fig.~\ref{fig1}, through the semitransparent mirror on the right. $\omega_l$ is close to the frequency $\omega_0$ of the center of the FPI mode spectrum. The FPI mode has Bose-operator ${{\hat{a}}}(t){{e}^{-i{{\omega }_{0}}t}}$ and is excited by the external field. Detuning $\delta ={{\omega }_0}-{{\omega }_l} \ll \omega_{0},\omega_{l}$. The small FPI in Fig.~\ref{fig1} has the size $\sim \lambda/2$, where $\lambda$ is the wavelength of the FPI mode. For certainty we suppose that the main mode of the small FPI is excited, so the FPI free spectral range is of the order of $\omega_0$ or $\omega_l$.  We assume the FPI cavity quality factor $Q\sim 10^3$; it can be achieved, for example, in  photonic crystal cavities   \ct{ZHANG2015374}. In the future, we want to satisfy   conditions for the dispersive bistability in the FPI with a nonlinear medium \ct{doi:10.1063/1.88632}, so we take $\kappa\leq \delta \leq $ few $\kappa$, where $\kappa$ is the half-width (HWHM) of the excited FPI mode. For such parameters we, with negligibly small error $\sim \delta/\kappa Q\ll 1$, neglect the excitation of all FPI modes, but the main FPI mode.    

\subsection{General formula for the photon number fluctuation spectrum in the cavity.}

We make Fourier-expansions of the amplitude operators of the field $\hat{a}(t)$ inside the FPI cavity and $\hat{a}_{in}(t)$ of the input field entering the cavity through the semitransparent mirror 
\beq
\hat{a}(t)= \frac{1}{\sqrt{2\pi }}\int\limits_{-\infty }^{\infty }{\hat{a}(\omega ){{e}^{-i\omega t}}d\omega }, \hspace{0.25cm} \text{and} \hspace{0.25cm}{{\hat{a}}_{in}}(t)=\frac{1}{\sqrt{2\pi }}\int\limits_{-\infty }^{\infty }{{{{\hat{a}}}_{in}}(\omega ){{e}^{-i\omega t}}d\omega }, \lb{FexAmp}
\eeq
where $\hat{a}(\omega )$ and $\hat{a}_{in}(\omega )$ are the  Fourier-component operators, $\omega = \omega_{opt}-\omega_l$ is the deviation of the field optical frequency 
 $\omega_{opt}$ from $\omega_l$. We  consider $T\left\langle :\hat{n}(t)\hat{n}(t'): \right\rangle$, where $\hat{n}$ is a photon number operator, $ :\hat{n}(t)\hat{n}(t'):$ means the normal ordering, $T$ is the time ordering and $\left\langle ... \right\rangle$ is the quantum averaging, see \ct{Mandel1995}, section 12.2.2. Using $\hat{n}(t)={{\hat{a}}^{+}}(t)\hat{a}(t)$ and the first of Fourier-expansions \rf{FexAmp} we write
\beq
%
T\left\langle :\hat{n}(t)\hat{n}(t'): \right\rangle =\frac{1}{{{\left( 2\pi  \right)}^{2}}}\int\limits_{-\infty }^{\infty }{d{{\omega }_{1}}d{{\omega }_{2}}d{{\omega }_{3}}d{{\omega }_{4}}\left\langle :{{{\hat{a}}}^{+}}(-{{\omega }_{1}})\hat{a}({{\omega }_{2}}){{{\hat{a}}}^{+}}(-{{\omega }_{3}})\hat{a}({{\omega }_{4}}): \right\rangle T\left[ {{e}^{-i({{\omega }_{1}}+{{\omega }_{2}})t-i({{\omega }_{3}}+{{\omega }_{4}})t'}} \right]}, \lb{autocorr_exp1}
\eeq
so we separate the time and the normal-ordering operations. 

Fourier-component operators of different frequencies commute, they are  uncorrelated with each other, so the mean $\left\langle {{{\hat{a}}}^{+}}(-{{\omega }_{1}})\hat{a}({{\omega }_{2}}){{{\hat{a}}}^{+}}(-{{\omega }_{3}})\hat{a}({{\omega }_{4}}) \right\rangle $ is not zero if ${{\omega }_{1}}=-{{\omega }_{2}}$ and ${{\omega }_{3}}=-{{\omega }_{4}}$, or ${{\omega }_{1}}=-{{\omega }_{4}}$ and ${{\omega }_{3}}=-{{\omega }_{2}}$. The commutator of the field Bose operators inside the cavity is
\beq
\left[ \hat{a}({{\omega }_{2}}),{{{\hat{a}}}^{+}}(-{{\omega }_{3}}) \right]=c({{\omega }_{2}})\delta ({{\omega }_{2}}+{{\omega }_{3}}), \lb{comm0}
\eeq
with  $(2\pi)^{-1}\int\limits_{-\infty }^{\infty }{c(\omega )d\omega =1}$ \ct{PhysRevA.30.1386}. Note that the cavity modifies the density of states of the quantum field, respectively to the free space \ct{PhysRev.69.674}, so commutation relations \rf{comm0} for Fourier-component operators in the cavity are different from commutation relations \rf{BCR_0} in the free space.  $c(\omega)$ for the FPI cavity is given by Eq.~\rf{comm_cav}.

Using \rf{comm0} we make the normal ordering in \rf{autocorr_exp1}   exchanging $\hat{a}({{\omega }_{2}})$ and ${{\hat{a}}^{+}}(-{{\omega }_{3}})$ 
\beq
  \left\langle :{{{\hat{a}}}^{+}}(-{{\omega }_{1}})\hat{a}({{\omega }_{2}}){{{\hat{a}}}^{+}}(-{{\omega }_{3}})\hat{a}({{\omega }_{4}}): \right\rangle = \lb{calc_01}\eeq\[ 
 n({{\omega }_{2}})n({{\omega }_{4}})\delta ({{\omega }_{1}}+{{\omega }_{2}})\delta ({{\omega }_{3}}+{{\omega }_{4}})+n({{\omega }_{2}})\left[ n({{\omega }_{4}})+c({{\omega }_{4}}) \right]\delta ({{\omega }_{1}}+{{\omega }_{4}})\delta ({{\omega }_{3}}+{{\omega }_{2}}). \]
We insert Eq.~\rf{calc_01} into Eq.~\rf{autocorr_exp1}, calculate the integral in  Eq.~\rf{autocorr_exp1} with $\delta$-functions and see that the integral from the first term in Eq.~\rf{calc_01} is $n^2$ so that
\beq
T\left\langle :\hat{n}(t)\hat{n}(t'): \right\rangle ={{n}^{2}}+\frac{1}{{{\left( 2\pi  \right)}^{2}}}\int\limits_{-\infty }^{\infty }{d{{\omega }_{2}}d{{\omega }_{4}}n({{\omega }_{2}})\left[ n({{\omega }_{4}})+c({{\omega }_{4}}) \right]T\left[ {{e}^{i({{\omega }_{4}}-{{\omega }_{2}})(t-t')}} \right]}.\lb{T-exp}
\eeq
The time-ordering operation in Eq.~\rf{T-exp} is
\beq
T\left[ {{e}^{i({{\omega }_{4}}-{{\omega }_{2}})(t-t')}} \right]=\left\{ \begin{matrix}
   {{e}^{i({{\omega }_{4}}-{{\omega }_{2}})(t-t')}},\ t\ge t'  \\
   {{e}^{i({{\omega }_{4}}-{{\omega }_{2}})(t'-t)}},\quad t<t'  \\
\end{matrix} \right.={{e}^{i({{\omega }_{4}}-{{\omega }_{2}})\left| t-t' \right|}}. \lb{Time_ord_0}
\eeq
Replacing in Eq.~\rf{T-exp} $\omega_2$ by a new variable $\omega ={{\omega }_{2}}-{{\omega }_{4}}$ and $\omega _4$ by $\omega'$  we find the second-order auto-correlation function ${{\delta }^{2}}n(\tau)$ for the photon number fluctuations
\beq
{{\delta }^{2}}n(\tau)\equiv T\left\langle :\hat{n}(t)\hat{n}(t'): \right\rangle -{{n}^{2}}=\frac{1}{{{\left( 2\pi  \right)}^{2}}}\int\limits_{-\infty }^{\infty }d\omega {d{{\omega }'}n(\omega +{{\omega }'})\left[ n({{\omega }'})+c({{\omega }'}) \right]{{e}^{-i\omega \left| \tau \right|}}}, \lb{Pn_numb_fl_corr}
\eeq
where $\tau = t-t'$. Wiener–Khinchin theorem (\ct{Champeney1987}, page 102) tells that the spectrum ${{\delta }^{2}}n\left( \omega  \right)$ of the photon number fluctuations is related to ${{\delta }^{2}}n(\tau)$ by the Fourier-transform 
\beq
{{\delta }^{2}}n(\tau )=(2\pi )^{-1}\int\limits_{-\infty }^{\infty }{{{\delta }^{2}}n\left( \omega  \right){{e}^{-i\omega \tau }}d\omega }, \lb{auto_corr_f}
\eeq
so we find $\delta^2n\left( \omega  \right)$ by the Fourier-transform of Eq.~\rf{Pn_numb_fl_corr}. In such a transform the integral over $d\tau$ is split  into two parts: from $-\infty$ to $0$ and from $0$ to $\infty$, taking into account the multiplier $\exp{(-i\omega|\tau|)}$. We carry out the Fourier-transform and come to
\beq
{{\delta }^{2}}n\left( \omega  \right)=\frac{1}{4\pi }\int\limits_{-\infty }^{\infty }{\left[ n(\omega '+\omega )+n(\omega '-\omega ) \right]\left[ n(\omega ')+c(\omega ') \right]d\omega }' \lb{spectrum_0}
\eeq
Making the replacement $\omega'\rightarrow \omega''=\omega'+\omega$ we see, that
%
%
$\int\limits_{-\infty }^{\infty }{n(\omega '+\omega )n(\omega ')d\omega }'=\int\limits_{-\infty }^{\infty }{n(\omega '')n(\omega ''-\omega )d\omega }''$,
%
%
so we re-write Eq.~\rf{spectrum_0} as
\beq
{{\delta }^{2}}n\left( \omega  \right)=\frac{1}{2\pi }\int\limits_{-\infty }^{\infty }{n(\omega +\omega ')n(\omega ')d\omega }'+\frac{1}{4\pi }\int\limits_{-\infty }^{\infty }{\left[ n(\omega '+\omega )+n(\omega '-\omega ) \right]c(\omega ')d\omega }'. \lb{fin_popfl_sp}
\eeq
The first term in Eq.~\rf{fin_popfl_sp} is the same as for the classical fluctuate field (\ct{Mandel1995}, section 9.8.3). One can obtain this term without the time and the normal orderings,  considering $\hat{a}$ in Eq.~\rf{autocorr_exp1} as a classical fluctuating variable. The second term in Eq.~\rf{fin_popfl_sp} appears due to quantum fluctuations applying the time and the normal orderings. Below we call the first term of \rf{fin_popfl_sp}  a classical contribution and the second one a quantum contribution to the photon number fluctuation spectrum. 

We see in Eq.~\rf{fin_popfl_sp} that ${{\delta }^{2}}n\left( \omega  \right)={{\delta }^{2}}n\left( -\omega  \right)$ as it must be for real $\delta^2n(\tau)$. The photon number variance ${{\delta }^{2}}n\equiv {(2\pi)^{-1} }\int\limits_{-\infty }^{\infty }{{{\delta }^{2}}n\left( \omega  \right)d\omega }=n\left( n+1 \right)$ corresponds to Bose-Einstein distribution. 

The meaning of $\omega$ in ${{\delta }^{2}}n\left( \omega  \right)$ 
is different from the meaning of $\omega$ in $n(\omega)$ or $\hat{a}(\omega)$. $\omega$ in ${{\delta }^{2}}n\left( \omega  \right)$ is the radio frequency. Otherwise, $\omega$ in $n(\omega)$ or $\hat{a}(\omega)$ is the deviation of the optical frequency $\omega_{opt}$ of the field inside the FPI from the central frequency $\omega_{l}$ of the input field spectra. 
\subsection{Photon number fluctuation spectra outside FPI: general formula}
We write the field power operator (in photons per second) in the free space ${{\hat{p}}_{\alpha }}(t)=\hat{a}_{\alpha }^{+}(t){{\hat{a}}_{\alpha }}(t)$ for the input $\alpha =in$, output $\alpha =out$ fields, and the field reflected from the input mirror of the interferometer $\alpha =out$. We will find 
\beq
%
T\left\langle :{{{\hat{p}}}_{\alpha }}(t){{{\hat{p}}}_{\alpha }}(t'): \right\rangle =\frac{1}{{{\left( 2\pi  \right)}^{2}}}\int\limits_{-\infty }^{\infty }{d{{\omega }_{1}}d{{\omega }_{2}}d{{\omega }_{3}}d{{\omega }_{4}}\left\langle :\hat{a}_{\alpha }^{+}(-{{\omega }_{1}}){{{\hat{a}}}_{\alpha }}({{\omega }_{2}})\hat{a}_{\alpha }^{+}(-{{\omega }_{3}}){{{\hat{a}}}_{\alpha }}({{\omega }_{4}}): \right\rangle T\left[ {{e}^{-i({{\omega }_{1}}+{{\omega }_{2}})t-i({{\omega }_{3}}+{{\omega }_{4}})t'}} \right]} \lb{outside_corr}
\eeq
We carry out the normal ordering in Eq.~\rf{outside_corr} using commutation relations for Bose operators in free space \ct{PhysRevA.30.1386}
\beq
\left[ {{{\hat{a}}}_{\alpha }}(\omega ),\hat{a}_{\alpha }^{+}(-\omega ') \right]=\delta (\omega +\omega ') \lb{BCR_0}
\eeq
so that
\[
\hspace{-2cm}\left\langle \hat{a}_{\alpha }^{+}(-{{\omega }_{1}}){{{\hat{a}}}_{\alpha }}({{\omega }_{2}})\hat{a}_{\alpha }^{+}(-{{\omega }_{3}}){{{\hat{a}}}_{\alpha }}({{\omega }_{4}}) \right\rangle =\left\langle \hat{a}_{\alpha }^{+}(-{{\omega }_{1}})\hat{a}_{\alpha }^{+}(-{{\omega }_{3}}){{{\hat{a}}}_{\alpha }}({{\omega }_{2}}){{{\hat{a}}}_{\alpha }}({{\omega }_{4}}) \right\rangle +\left\langle \hat{a}_{\alpha }^{+}(-{{\omega }_{1}}){{{\hat{a}}}_{\alpha }}({{\omega }_{4}}) \right\rangle \delta ({{\omega }_{2}}+{{\omega }_{3}}).
\]
Similar to the case of Eq.~\rf{autocorr_exp1}, we note that operators of different frequencies commute and do not correlate with each other, therefore
\[
  \hspace{-2cm}\left\langle \hat{a}_{\alpha }^{+}(-{{\omega }_{1}}){{{\hat{a}}}_{\alpha }}({{\omega }_{2}})\hat{a}_{\alpha }^{+}(-{{\omega }_{3}}){{{\hat{a}}}_{\alpha }}({{\omega }_{4}}) \right\rangle =\left\langle \hat{a}_{\alpha }^{+}({{\omega }_{2}}){{{\hat{a}}}_{\alpha }}({{\omega }_{2}}) \right\rangle \left\langle \hat{a}_{\alpha }^{+}({{\omega }_{4}}){{{\hat{a}}}_{\alpha }}({{\omega }_{4}}) \right\rangle \delta ({{\omega }_{1}}+{{\omega }_{2}})\delta ({{\omega }_{3}}+{{\omega }_{4}})+ \]\beq 
 \left\langle \hat{a}_{\alpha }^{+}({{\omega }_{4}}){{{\hat{a}}}_{\alpha }}({{\omega }_{4}}) \right\rangle \left[ \left\langle \hat{a}_{\alpha }^{+}({{\omega }_{2}}){{{\hat{a}}}_{\alpha }}({{\omega }_{2}}) \right\rangle +1 \right]\delta ({{\omega }_{1}}+{{\omega }_{2}})\delta ({{\omega }_{3}}+{{\omega }_{4}}) \lb{open_sp2}
\eeq
Inserting Eq.~\rf{open_sp2} into the integral in Eq.~\rf{outside_corr}, we see that the first term in Eq.~\rf{open_sp2} is the mean power square ${{\left\langle {{{\hat{p}}}_{\alpha }}(t) \right\rangle }^{2}}\equiv p_{\alpha }^{2}$. Taking the integral in Eq.~\rf{outside_corr} over $d{{\omega }_{1}}$ and $d{{\omega }_{3}}$, inserting there the power spectra ${{p}_{\alpha }}({{\omega }_{2,4}})=\left\langle \hat{a}_{\alpha }^{+}({{\omega }_{2,4}}){{{\hat{a}}}_{\alpha }}({{\omega }_{2,4}}) \right\rangle $ we come to the  auto-correlation function for the field power fluctuations in free space 
\beq
{{\delta }^{2}}{{p}_{\alpha }}(\tau)\equiv\left\langle \hat{p}_{\alpha }(t)\hat{p}_{\alpha }(t') \right\rangle -p_{\alpha }^2=\frac{1}{\left( 2\pi  \right)^2}\int\limits_{-\infty }^{\infty }d\omega'd\omega p_{\alpha }(\omega'-\omega )\left[ p_{\alpha }(\omega')+1 \right]e^{-i\omega \left| \tau \right|}, \lb{auto_fsp_0}
\eeq
where $\tau=t-t'$ and we replace  $\omega_4$ by $\omega ={{\omega }_{2}}-{{\omega }_{4}}$; $\omega_2$ by $\omega'$. The integral over $d{{\omega }_{2}}$  in the second term in Eq.~\rf{auto_fsp_0} leads to $(2\pi)^{-1}\int\limits_{-\infty }^{\infty }{d{{\omega }_{2}}{{p}_{\alpha }}({{\omega }_{2}}-\omega )}={{p}_{\alpha }}$, the integral over $d\omega$ in this term gives $(2\pi)^{-1}\int\limits_{-\infty }^{\infty }{{{e}^{-i\omega \left| \tau \right|}}d\omega }=\delta (\tau)$, so the auto-correlation function is 
\beq
{{\delta }^{2}}{{p}_{\alpha }}(\tau)=\frac{1}{{{\left( 2\pi  \right)}^{2}}}\int\limits_{-\infty }^{\infty }{d\omega '{{p}_{\alpha }}(\omega '-\omega ){{p}_{\alpha }}(\omega '){{e}^{-i\omega \left| \tau\right|}}d\omega }+{{p}_{\alpha }}\delta (\tau). \lb{autocorr_fr_sp_22}
\eeq
The Fourier-transform of Eq.~\rf{autocorr_fr_sp_22} leads to the spectrum ${{\delta }^{2}}{{p}_{\alpha }}(\omega )$ of the field power fluctuations in the free space
\beq
{{\delta }^{2}}{{p}_{\alpha }}(\omega )=\frac{1}{4\pi }\int\limits_{-\infty }^{\infty }{\left[ {{p}_{\alpha }}(\omega '-\omega ){{p}_{\alpha }}(\omega ')+{{p}_{\alpha }}(\omega '+\omega ){{p}_{\alpha }}(\omega ') \right]d\omega '}+{{p}_{\alpha }}. \lb{pfs_0}
\eeq
Since $\int\limits_{-\infty }^{\infty }{d\omega '{{p}_{\alpha }}(\omega '-\omega ){{p}_{\alpha }}(\omega ')}=\int\limits_{-\infty }^{\infty }{d\omega '{{p}_{\alpha }}(\omega '+\omega ){{p}_{\alpha }}(\omega ')}$ we write Eq.~\rf{pfs_0} as 
\beq
{{\delta }^{2}}{{p}_{\alpha }}(\omega )=\frac{1}{2\pi }\int\limits_{-\infty }^{\infty }{{{p}_{\alpha }}(\omega '-\omega ){{p}_{\alpha }}(\omega ')d\omega '}+{{p}_{\alpha }}. \lb{free_sp_final}
\eeq
The first term in Eq.~\rf{free_sp_final} is a "colored" part of the spectrum. The second term is a white noise with constant spectrum power density $p_{\alpha}$.  The first term in Eq.~\rf{free_sp_final} is the classical contribution, similar to the first term in Eq.~\rf{fin_popfl_sp}.  The second one is the quantum contribution. 

Eq.~\rf{free_sp_final} is different from the result for the classical field (\ct{Mandel1995}, section 9.8.3) in the second term $p_{\alpha}$, which is the spectral density of the "white" quantum noise. Note the difference between the quantum (the second) terms in Eqs.~\rf{fin_popfl_sp}  and \rf{free_sp_final}. The quantum term in Eq.~\rf{fin_popfl_sp} is a convolution of  "colored" quantum noise with the field spectrum in the cavity, while there is  a "white" quantum noise term in Eq.~\rf{free_sp_final} for the free space.

Dimentionalities of $n(\omega )$ and ${{\delta }^{2}}n(\omega )$ are the photon number per Hz and the photon number square per Hz, respectively. Dimentionalities of ${{p}_{\alpha }}(\omega )$   (${{\delta }^{2}}{{p}_{\alpha }}(\omega )$) are  the photon number (the photon number square) per second per Hz. 
\subsection{The model of the FPI with quantum input field}
\begin{figure}[thb]
\includegraphics[width=12 cm]{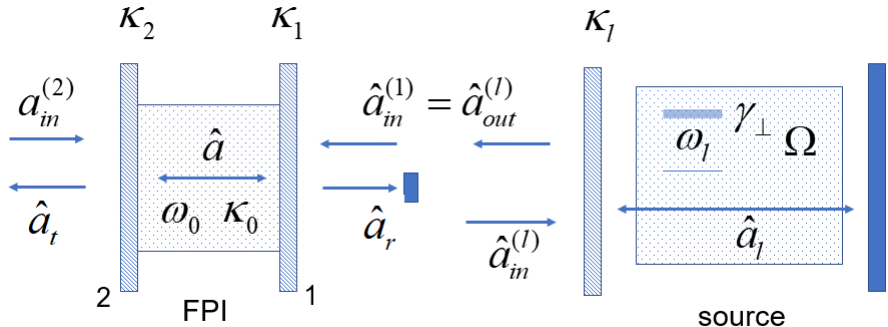}
\caption{FPI (on the left) with the source of the quantum field (on the right). The field $\hat{a}_r$, reflected from the FPI input mirror 1, is isolated from the source. FPI parameters explained in the main text, and parameters of the field source described in the Appendix.}\label{Fig2}
\end{figure}

We consider the field and the photon number fluctuation spectra inside and outside the small FPI; the main FPI cavity mode is excited by the  quasi-monochromatic quantum field taken, for example, from a LED or a laser. Fig.~\ref{Fig2} shows the scheme of the FPI (on the left) and the input field source (on the right).  
%
%
%
%
Bose-operator $\hat{a}$ of the FPI mode amplitude satisfies the equation
\beq
\dot{\hat{a}}=-\left( i\delta +{{\kappa }_{t}} \right)\hat{a}+\sqrt{2{{\kappa }_{0}}}\hat{a}_{in}^{(0)}+\sqrt{2{{\kappa }_{1}}}\hat{a}_{in}^{(1)}+\sqrt{2{{\kappa }_{2}}}\hat{a}_{in}^{(2)},\lb{eq_FPI}
\eeq
written  with the help of the input-output theory \ct{PhysRevA.30.1386}. In Eq.~\rf{eq_FPI}, ${{\kappa }_{t}}={{\kappa }_{1}}+{{\kappa }_{2}}+{{\kappa }_{0}}$ is the decay rate of the mode due to the field escape through the FPI semitransparent mirrors 1 and 2 with rates $\kappa_{1,2}$ and the absorption inside the FPI with the rate $\kappa_0$. Bose-operators $\hat{a}_{in}^{(0)}$ and $\hat{a}_{in}^{(2)}$ corresponding to zero temperature baths, uncorrelated with each other and  related with the absorption and the field escape through the FPI mirror 2. The input field with the Bose-operator $\hat{a}_{in}^{(1)}$ is the output $\hat{a}_{out}^{(l)}$ of the source in Fig.~\ref{Fig2}. 

Making the Fourier transform in Eq.~\rf{eq_FPI} and solving the equation for Fourier-component operators, we find the Fourier-component operator of the FPI mode
\beq
\hat{a}(\omega )=\frac{\sqrt{2{{\kappa }_{0}}}\hat{a}_{in}^{(0)}(\omega )+\sqrt{2{{\kappa }_{1}}}\hat{a}_{in}^{(1)}(\omega )+\sqrt{2{{\kappa }_{2}}}\hat{a}_{in}^{(2)}(\omega )}{{{\kappa }_{t}}+i\left( \delta -\omega  \right)}, \lb{FPI_fourier}
\eeq
where $\omega ={{\omega }_{opt}}-{{\omega }_{0}}$, ${\omega }_{opt}$ is the optical frequency of the mode, $\hat{a}_{in}^{(0,2)}(\omega)$ and $\hat{a}_{in}^{(1)}(\omega)$ are Fourier-component operators of baths and the input field. The baths and the input field operators obey the free space Bose-commutation relations \ct{PhysRevA.30.1386}
\beq
        [\hat{a}_{in}^{(\alpha)}(\omega),\hat{a}_{in}^{+(\alpha)}(\omega')] = \delta(\omega+\omega'), \lb{BCR}
\eeq
$\alpha=0,1,2$. Substituting the expression \rf{FPI_fourier} into the relation $\left\langle {{{\hat{a}}}^{+}}(-\omega )\hat{a}(\omega ') \right\rangle =n(\omega )\delta (\omega +\omega ')$ we obtain the spectrum $n(\omega )$ of the  FPI mode
\beq
n(\omega )=\frac{{{\kappa }_{1}}}{{{\kappa }_{t}}}{{p}_{in}}(\omega )L(\delta -\omega ,{{\kappa }_{t}}),\lb{FPI_field_sp}
\eeq
where ${{p}_{in}}(\omega )$ is the input field spectrum satisfying  $\left\langle \hat{a}_{in}^{(1)+}(-\omega )\hat{a}_{in}^{(1)}(\omega ') \right\rangle ={{p}_{in}}(\omega )\delta (\omega +\omega ')$. In Eq.~\rf{FPI_field_sp} and below we denote the Lorenz spectrum
\beq
L(\omega ,\kappa )=\frac{2\kappa }{{{\omega }^{2}}+{{\kappa }^{2}}}, \hspace{0.5cm} {{\left( 2\pi  \right)}^{-1}}\int\limits_{-\infty }^{\infty }{L(\omega ,\kappa )d\omega }=1. \lb{Lorenz}
\eeq
Using Eq.~\rf{FPI_fourier} and Bose-commutation relations \rf{BCR}, we find 
\beq
[\hat{a}(\omega '),{{\hat{a}}^{+}}(-\omega )]=L(\delta -\omega ,{{\kappa }_{t}})\delta (\omega +\omega ') \equiv c(\omega)\delta (\omega +\omega '),\lb{comm_cav}
\eeq
where $c(\omega)$ is a "commutator spectrum", see Eq.~\rf{comm0}.

The input field spectrum  is
\beq
{{p}_{in}}(\omega )={{p}_{in}}L(\omega ,{{\gamma }_{l}}), \hspace{0.5cm} {{\gamma }_{l}}=\frac{{{\gamma }_{\max }}}{1+{{p}_{in}}/{{\kappa }_{l}}},\lb{inp_f_sp}
\eeq
where ${p}_{in}$ is the input power in photons per second, $\gamma_l$ is the spectrum HWHM with the maximum value $\gamma_{max}$; $\kappa_l$ is the  input field emission rate from the mirror of the source in Fig.~\ref{Fig2}. When  $p_{in}$ increases, $\gamma_l$ decreases. We identify the LED, the intermediate, and the lasing regime when $\gamma_l \gg \kappa_l$, $\gamma_l \sim \kappa_l$, and $\gamma_l \ll \kappa_l$, correspondingly, in the source. The derivation of expressions \rf{inp_f_sp} is in the Appendix.
\subsection{Explicit expressions for spectra inside FPI}
The field spectrum $n(\omega)$ inside FPI is given by Eqs.~\rf{FPI_field_sp} and \rf{inp_f_sp}.  Using\\
%
%
$(2\pi )^{-1}\int\limits_{-\infty }^{\infty }{L(\omega ,{{\gamma }_{1}})L(\delta -\omega ,{{\gamma}_{2}})d\omega }=L(\delta ,{{\gamma }_{1}}+{{\gamma }_{2}})$ 
%
%
we calculate the integral\\  $n=(2\pi )^{-1}\int\limits_{-\infty }^{\infty }{n(\omega )d\omega }$ and find the mean photon number in the FPI mode  
\beq
n=\frac{{{\kappa }_{1}}}{{{\kappa }_{t}}}{{p}_{in}}L(\delta ,{{\kappa }_{t}}+{{\gamma }_{l}}).\lb{M_Pn_N}
\eeq
With $n(\omega)$ from Eq.~\rf{FPI_field_sp} and $c(\omega)$ from Eq.~\rf{comm_cav} we obtain, from Eq.~\rf{fin_popfl_sp}, the FPI mode photon number (or the field power) fluctuation spectrum
\beq
{{\delta }^{2}}n\left( \omega  \right)={{\left( \frac{{{p}_{in}}{{\kappa }_{1}}}{{{\kappa }_{t}}} \right)}^{2}}{{J}_{0}}(\omega )+\frac{{{p}_{in}}{{\kappa }_{1}}}{{{\kappa }_{t}}}{{J}_{1}}(\omega ), \lb{Phot_num_fl_ins}
\eeq
where
\beq
{{J}_{0}}(\omega )=\frac{1}{2\pi }\int\limits_{-\infty }^{\infty }{L(\omega '-\omega ,{{\gamma }_{l}})L(\omega '-\omega -\delta ,{{\kappa }_{t}})L(\omega ',{{\gamma }_{l}})L(\omega '-\delta ,{{\kappa }_{t}})d\omega '},\lb{J0}
\eeq\beq
\hspace{-3cm}{{J}_{1}}(\omega )=\frac{1}{4\pi }\int\limits_{-\infty }^{\infty }{\left[ L(\omega '-\omega ,{{\gamma }_{l}})L(\omega '-\omega -\delta ,{{\kappa }_{t}})+L(\omega '+\omega ,{{\gamma }_{l}})L(\omega '+\omega -\delta ,{{\kappa }_{t}}) \right]L(\omega '-\delta ,{{\kappa }_{t}})d\omega '}\lb{J1}
\eeq
One can find some cumbersome explicit expressions for  $J_{0,1}(\omega)$. We do not write them
here.
\subsection{Explicit spectra outside FPI}
\subsubsection{The field spectra}
Spectra  of the transmitted $p_{t}(\omega)$, the absorbed $p_{0}(\omega)$ fields  and the transmitted $p_t$ and the absorbed $p_0$ powers are
\beq
p_{t,0}(\omega)=2\kappa_{2,0}n(\omega), \hspace{0.5cm} p_{t,0}=2\kappa_{2,0}n.\lb{TA_FS}
\eeq
The reflected field  is ${{\hat{a}}_{r}}=\sqrt{2{{\kappa }_{1}}}\hat{a}-\hat{a}_{in}^{(1)}$, according to the boundary conditions on the FPI input mirror 1 in Fig.~\ref{Fig2}. Taking $\hat{a}(\omega)$ from Eq.~\rf{FPI_fourier}, we see that the reflected field Fourier component is
\beq
{{\hat{a}}_{r}}(\omega )=2\sqrt{{{\kappa }_{1}}}\frac{\sqrt{{{\kappa }_{0}}}\hat{a}_{in}^{(0)}(\omega )+\sqrt{{{\kappa }_{1}}}\hat{a}_{in}^{(1)}(\omega )+\sqrt{{{\kappa }_{2}}}\hat{a}_{in}^{(2)}(\omega )}{{{\kappa }_{t}}+i\left( \delta -\omega  \right)}-\hat{a}_{in}^{(1)}(\omega )\lb{r_field}
\eeq
Inserting Eq.~\rf{r_field} into $\left\langle \hat{a}_{r}^{+}(-\omega ){{{\hat{a}}}_{r}}(\omega ') \right\rangle ={{p}_{r}}(\omega )\delta (\omega +\omega ')$, taking into account that only the input field $\hat{a}_{in}^{(1)}(\omega )$ gives a non-zero contribution to ${{p}_{r}}(\omega )$, calculating the mean values and expressing the result in terms of Lorenz spectra \rf{Lorenz}, we obtain the reflected field spectrum
\beq
{{p}_{r}}(\omega )={{p}_{in}}\left[ 1-\frac{2{{\kappa }_{1}}({{\kappa }_{2}}+{{\kappa }_{0}})}{{{\kappa }_{t}}}L(\omega -\delta ,{{\kappa }_{t}}) \right]L(\omega ,{{\gamma }_{l}}).\lb{RFS_Lor}
\eeq
One can see that ${{p}_{r}}(\omega )+{{p}_{t}}(\omega )+{{p}_{0}}(\omega )={{p}_{in}}(\omega )$, as it must be.
\subsubsection{The field power fluctuation spectra}
We substitute $p_t(\omega)$ and $p_t$ from Eqs.~\rf{TA_FS}  into Eq.~\rf{free_sp_final}, carry out the integration, and find the power fluctuation spectrum of the transmitted field
\beq
{{\delta }^{2}}{{p}_{t}}(\omega )=2{{\kappa }_{2}}\left[ {{\left( \frac{{{\kappa }_{1}}}{{{\kappa }_{t}}}{{p}_{in}} \right)}^{2}}{{J}_{0}}(\omega )+n \right], \lb{thans_f_sp}
\eeq
where ${J}_{0}(\omega)$ is given by Eq.~\rf{J0}. 

We insert ${{p}_{r}}(\omega )$ from Eq.~\rf{RFS_Lor} and $p_r$ from Eq.~\rf{TA_FS} into Eq.~\rf{free_sp_final}, calculate integrals, and find the spectrum of the reflected field power fluctuations
\beq
{{\delta }^{2}}{{p}_{r}}(\omega )=p_{in}^{2}\left\{ L(\omega ,2{{\gamma }_{l}})-\frac{2{{\kappa }_{1}}({{\kappa }_{2}}+{{\kappa }_{0}})}{{{\kappa }_{t}}}{{J}_{2}}(\omega )+{{\left( \frac{2{{\kappa }_{1}}({{\kappa }_{2}}+{{\kappa }_{0}})}{{{\kappa }_{t}}} \right)}^{2}}{{J}_{0}}(\omega ) \right\}+{{p}_{r}},\label{RFPF}
\eeq
where
\beq
{{J}_{2}}(\omega )=\frac{1}{2\pi }\int\limits_{-\infty }^{\infty }{L(\omega '-\omega ,{{\gamma }_{l}})L(\omega ',{{\gamma }_{l}})\left[ L(\omega '-\omega -\delta ,{{\kappa }_{t}})+L(\omega '-\delta ,{{\kappa }_{t}}) \right]d\omega '}\lb{J2}
\eeq
and $J_0(\omega)$ is given by Eq.~\rf{J0}.
\subsubsection{Reflection and transmission coefficients}
We calculate the reflected field power $p_r$  integrating Eq.~\rf{RFS_Lor}  and find the coefficient $R$ of the reflection from the FPI input mirror 1 in Fig.~\ref{Fig2}
\beq
R\equiv \frac{{{p}_{r}}}{{{p}_{in}}}=1-\frac{2{{\kappa }_{1}}({{\kappa }_{2}}+{{\kappa }_{0}})}{{{\kappa }_{t}}}L(\delta ,{{\kappa }_{t}}+{{\gamma }_{l}}).\lb{R_coef}
\eeq
Taking $p_t=2\kappa_2n$ and the expression \rf{M_Pn_N} for $n$, we find the FPI transmission coefficient
\beq
T\equiv \frac{{{p}_{t}}}{{{p}_{in}}}=\frac{2{{\kappa }_{1}}{{\kappa }_{2}}}{{{\kappa }_{t}}}L(\delta ,{{\kappa }_{t}}+{{\gamma }_{l}}).\lb{T_coef}
\eeq
When $\kappa_0=0$, the absorption inside the FPI is absent, then $T+R=1$. When $\gamma_l\rightarrow 0$, Eqs.~\rf{R_coef} and \rf{T_coef} come to  expressions for the $T$ and $R$ for the FPI with the  monochromatic input  \ct{Akhmanov1997}.
\section{Results}\label{sec3}
We demonstrate examples of spectra of the FPI  with quantum input at the different  input field powers. Suppose the FPI input is  taken from the small field source as  the quantum dot LED or laser with the photonic crystal cavity \ct{ZHANG2015374}. The source cavity quality factor is $Q=10^3$;  the frequency of the center of the input field spectrum corresponds to the wavelength $\lambda_l=1.55$~$\mu$m. Such a field source  with the mean cavity photon number $n_l=1$ produces about $8\cdot {{10}^{11}}$ photons per second ($\approx 0.07$ $\mu$Wt). We take the rate ${{\kappa }_{l}}=4\cdot {{10}^{11}}$~rad/sec  of the field escape from the source cavity as a normalizing factor.  

We will see interesting features of FPI spectra by changing the normalized FPI input power $p_{in}/\kappa_l = 2n_l$ in the range from $0$ to $50$. $n_l$ in the range $0<n_l\leq 25$   corresponds to the LED ($n_l<1$), the intermediate ($n_l\sim 1$), and the lasing ($n_l\gg 1$) radiation from the source. $n_l\leq 25$  obtained in the  photonic crystal laser with 100 -- 1000 resonant emitters (quantum dots) with the small lasing mode volume $V$, for example,  $V\sim 10{{\left( {{\lambda }_{l}}/2{{n}_{r}} \right)}^{3}}$, where ${{\left( {{\lambda }_{l}}/2{{n}_{r}} \right)}^{3}}$ is the minimum cavity mode volume, $n_r$ is the refractive index inside the source cavity. 

We take $\gamma_{max} = 3\kappa_l$ in Eq.~\rf{inp_f_sp}; see the expression for $\gamma_{max}$ in the Appendix. We suppose that the FPI is from the photonic crystal, similar to the cavity of the input field source, so we  take the  decay rates in the FPI cavity of the order of $\kappa_l$, namely ${{\kappa }_{1}}/{{\kappa }_{l}}={{\kappa }_{2}}/{{\kappa }_{l}}=0.5$, ${{\kappa }_{0}}/{{\kappa }_{l}}=0.1$. We set the detuning $\delta /{{\kappa }_{l}}=5$, which is relatively large respectively to the FPI mode total decay rate $\kappa_t/\kappa_l = 1.1$. Relatively large detuning is necessary for the optical bistability in the small FPI with a quantum field and nonlinear medium \ct{doi:10.1080/00107518308210690, PhysRevA.19.2074}, which we will consider in the future. The chosen $\delta$ is not too large, so the main FPI mode is  effectively excited by the input field, while the excitation of the other FPI modes is negligibly small.     

When  $\delta/\kappa_l=5$, the mean FPI cavity photon number $n\sim 1$,  even for high power of the input field  $p_{in}/\kappa_l = 50$, when the input is practically coherent radiation of a narrow spectrum, with a small HWHM  $\gamma_l/\kappa_l \ll 1$, see Fig.~\ref{fig3}a. Such a small number of photons in the FPI cavity confirms the importance of  quantum analysis.
\subsection{Reflection and transmission coefficients}
\begin{figure}[thb]
\includegraphics[width=7.0 cm]{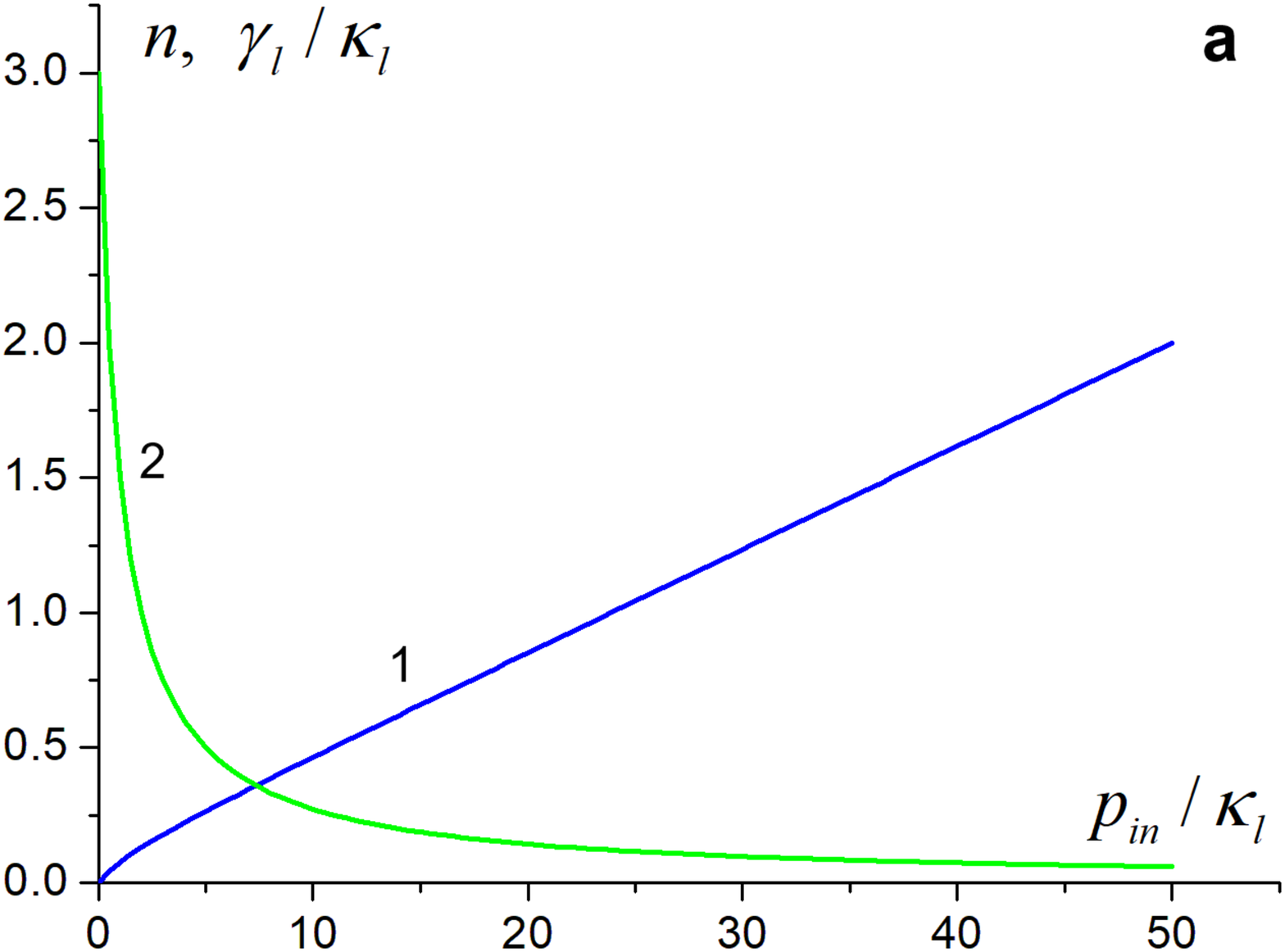}\hspace{0.5cm}\includegraphics[width=6.5 cm]{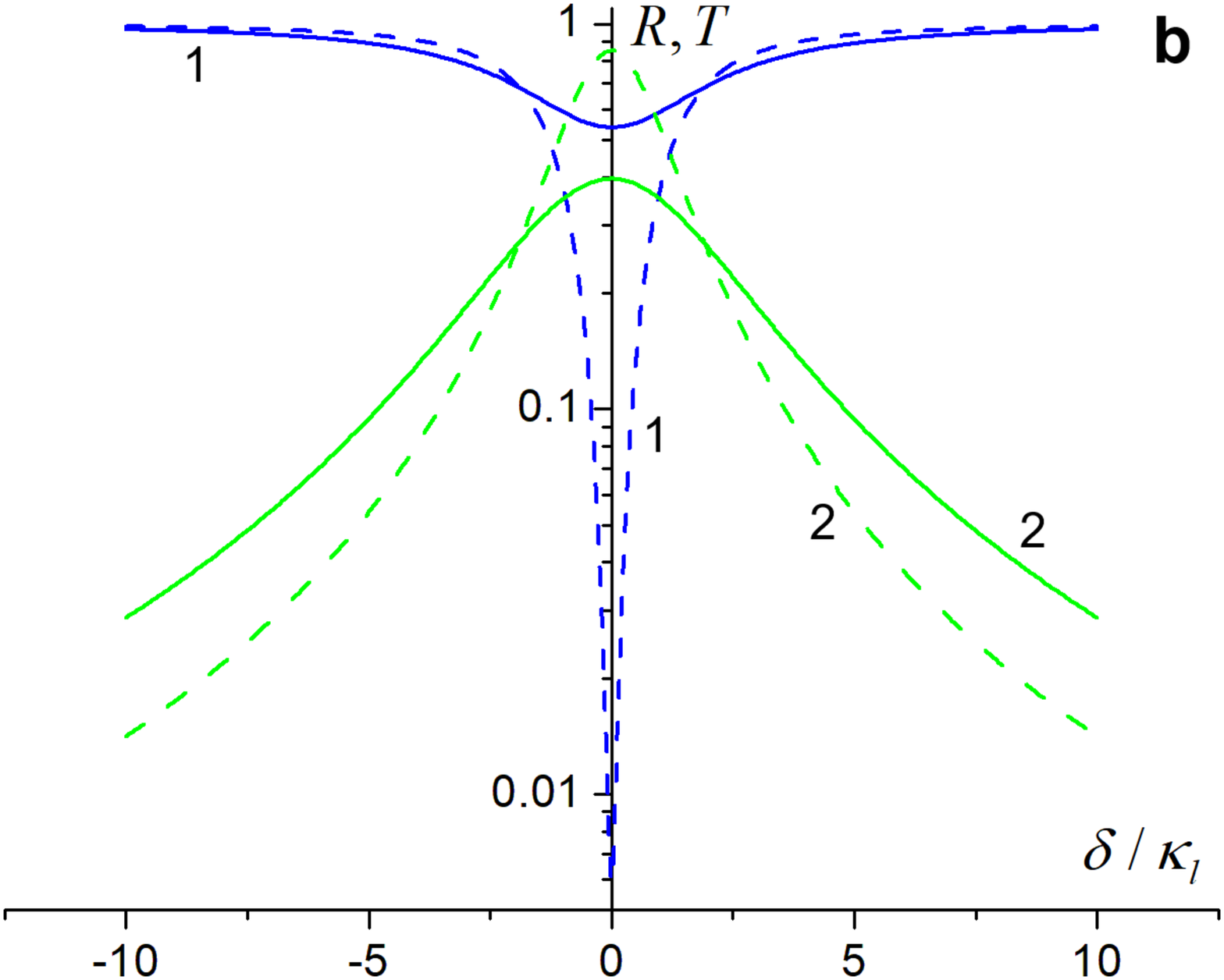}
\caption{(a) The mean photon number $n$ in the FPI mode  (curve 1) and the HWHM $\gamma_l$ of the input field spectrum (curve 2) versus the power $p_{in}$ of the input field for the detuning $\delta/\kappa_l=5$. Note that $n\sim 1$, at a high input power $p_{in}/\kappa_l\gg 1$. (b) Reflection $R$ (curves 1) and transmission $T$ (curves 2) coefficients of the FPI versus the detuning $\delta$. Solid curves are for the input field spectrum of a finite HWHM $\gamma_l/\kappa_l = 1.5$ (at $p_{in}/\kappa_l = 1$), dashed curves are for the monochromatic input with $\gamma_l/\kappa_l \rightarrow 0$, values of other parameters given in the text. Near $\delta/\kappa_l=0$, the transmittance for the finite spectral  width field is lower (and the reflection is higher) than for the monochromatic field -- compare the solid and the dashed curves with the same numbers.}\label{fig3}
\end{figure}

The reflection $R(\delta)$ and the transmission $T(\delta)$ coefficients characterize the FPI. Eqs.~\rf{R_coef} dive $R(\delta)$, and $T(\delta)$, shown on Fig.~\ref{fig3}~b, for the FPI with quantum input. $R(\delta)$ and $T(\delta)$  are symmetric relative to $\delta=0$. We see from Fig.~\ref{fig3}b that the FPI has a lower transmission and higher reflection for the finite spectral width input  than for the monochromatic input field. Indeed, the non-monochromatic input field includes the frequency components detuned from the exact resonance with the FPI mode even at $\delta=0$. 
%
%
%
%
%
\subsection{Field spectra}
The field spectra outside the FPI have the spectrum power densities ${{p}_{\alpha }}({{\omega }_{opt}}-{{\omega }_{l}})$ (in photons per second per Hz): transmitted $\alpha =t$, see Eq.~\rf{TA_FS}; input $\alpha =in$, Eq.~\rf{inp_f_sp}; and reflected from the mirror of the FPI, $\alpha =r$, Eq.~\rf{RFS_Lor}. The physical meaning of, for example, ${{p}_{r}}(\omega )d\omega /{{p}_{in}}$ is the part of the input field power in a narrow frequency interval ${{\omega }_{l}}+\omega \div {{\omega }_{l}}+\omega +d\omega$ reflected from the FPI input mirror. 
%
%
\begin{figure}[thb]
\includegraphics[width=7.0 cm]{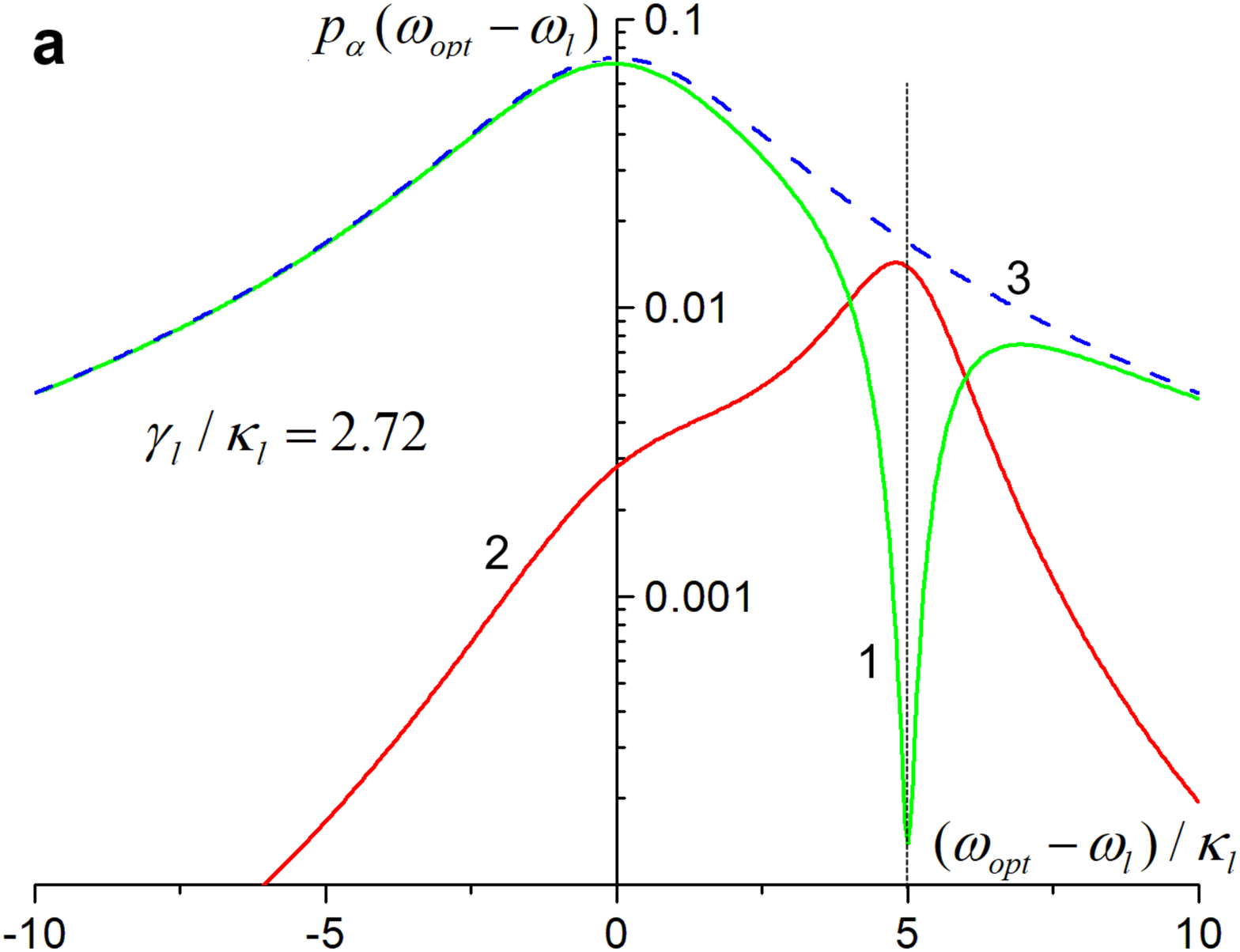}\hspace{0.5cm}\includegraphics[width=7.0 cm]{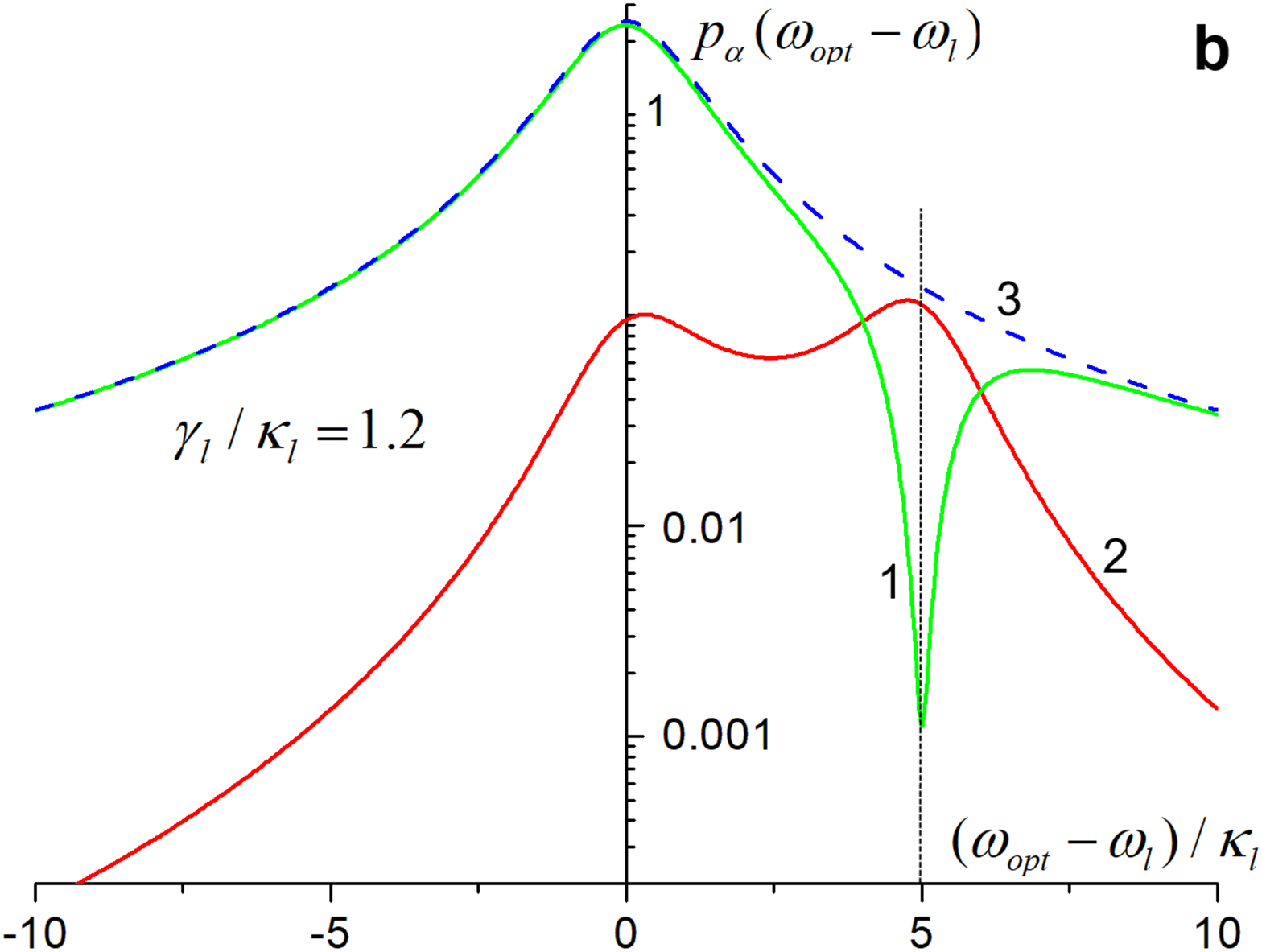}\\
\includegraphics[width=7.0 cm]{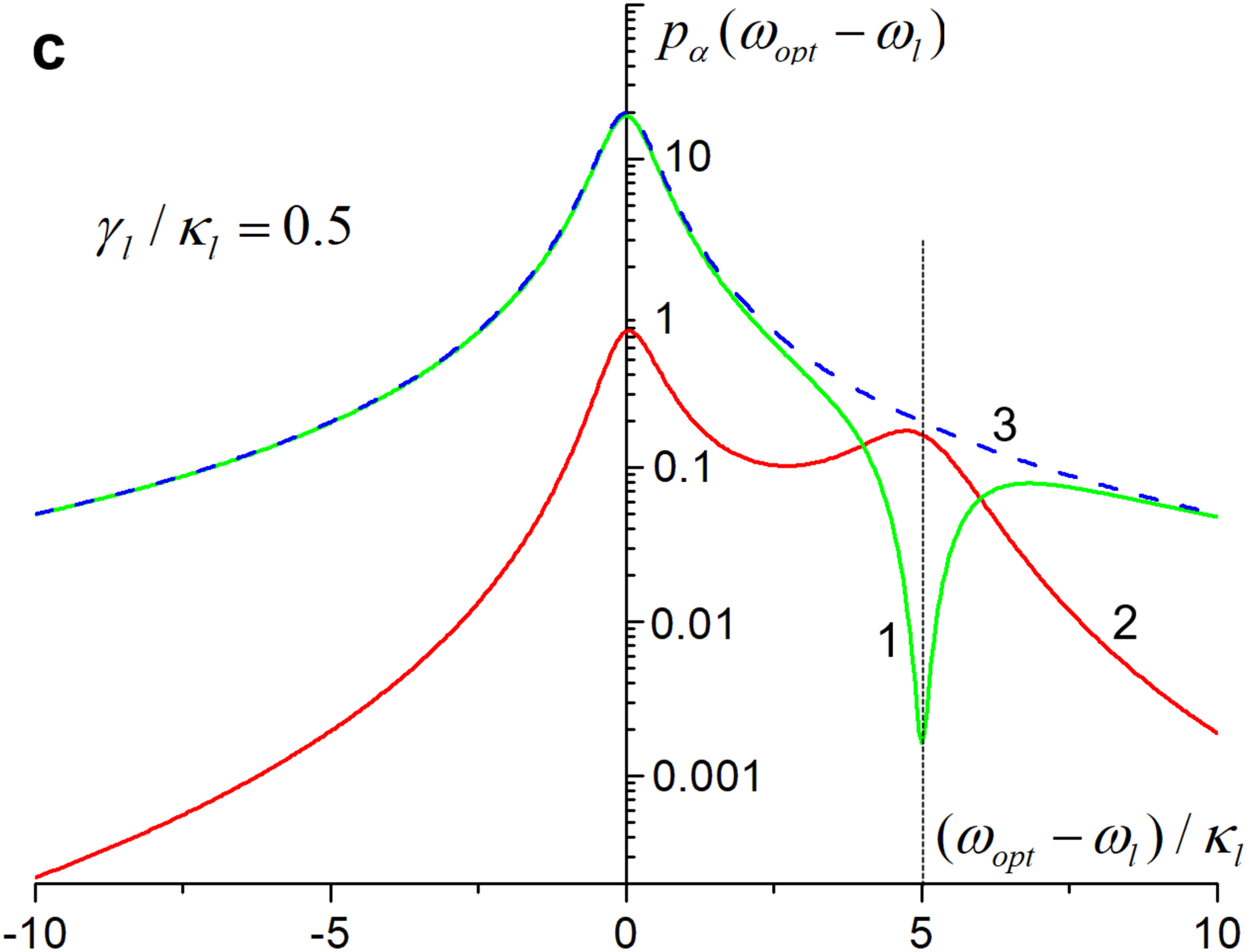}\hspace{0.5cm}\includegraphics[width=7.0 cm]{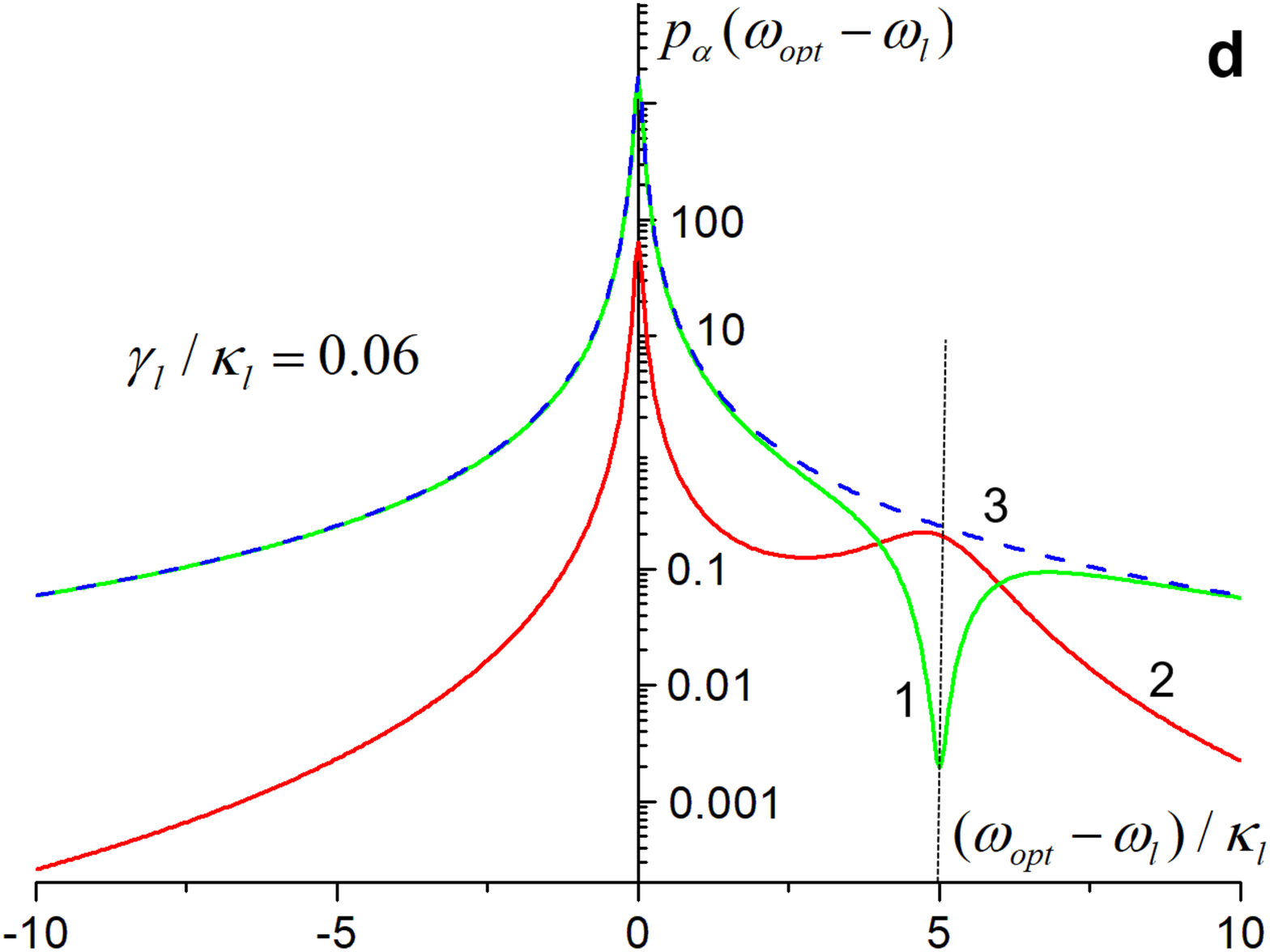}
\caption{Spectral power densities ${{p}_{\alpha }}({{\omega }_{opt}}-{{\omega }_{l}})$ for reflected (curves 1), transmitted (2), and input (3) fields. The power $p_{in}$ increases from (a) to (d): ${{p}_{in}}/{{\kappa }_{l}}=0.1$ (a), 1.5 (b), 5 (c), and 50 (d). The input field spectrum HWHM $\gamma_l$ decreases with the $p_{in}$; $\gamma_l/\kappa_l$ values are in the figures. The minima of the transmitted and local maxima of the reflected field spectra  correspond to the center of the FPI mode marked by the vertical dashed line. Maxima at ${{\omega }_{opt}}={{\omega }_{l}}$ (the horizontal axis zero) correspond to the center of the input field spectrum.}
\label{fig4}
\end{figure}   
%
%
Figs.~\ref{fig4}a-d show ${{p}_{\alpha }}({{\omega }_{opt}}-{{\omega }_{l}})$. The input field power $p_{in}$ increased, and the input field spectra HWHM $\gamma_l$  decreased from Fig.~\ref{fig4}a to Fig.~\ref{fig4}d. 
The  reflected field spectrum ${{p}_{r}}({{\omega }_{opt}}-{{\omega }_{l}})$ (green curves 1) has the same structure at any $p_{in}$ and $\gamma_l$.  There is the peak  at the center of the input field spectrum at ${{\omega }_{opt}}={{\omega }_{l}}$ and the gap at the center of the FPI mode spectrum at ${{\omega }_{opt}}={{\omega }_{0}}$. The peak is narrower and higher, and the gap has a smaller depth   at the higher $p_{in}$ and smaller $\gamma_l$; see the green curves 1. 

The structure of the transmitted field spectrum (the red curves 2) changed with $p_{in}$ and $\gamma_l$. For Fig.~\ref{fig4}a, the FPI mode is excited by a weak broadband input field with ${{p}_{in}}/{{\kappa }_{l}}=0.1$ and ${{\gamma }_{l}}/{{\kappa }_{l}}=2.72>{{\kappa }_{t}}/{{\kappa }_{l}}=1.1$, so the input field spectrum is broader than the empty FPI mode spectrum. The transmitted field spectrum in Fig.~\ref{fig4}a is broad and asymmetric,   
with a single maximum at the center ${{\omega }_{opt}}={{\omega }_{0}}$ of the FPI mode.

When we go from Fig.~\ref{fig4}a to Fig.~\ref{fig4}d, the input power $p_{in}$ grows with the narrowing of the input field spectrum. For Fig.~\ref{fig4}b  ${{p}_{in}}/{{\kappa }_{l}}=1.5$, and ${{\gamma }_{l}}/{{\kappa }_{l}}=1.2$ -- close to the HWHM ${{\kappa }_{t}}/{{\kappa }_{l}}=1.1$ of the empty FPI mode. The peak at ${{\omega }_{opt}}-{{\omega }_{l}}=\delta$ appears in the transmitted field spectrum in Fig.~\ref{fig4}b. Parameters of Fig.~\ref{fig4}b are such that both maxima in the transmitted field spectra have approximately the same height. The maxima slightly shifted toward each other relative to the maxima of the input field spectra (at zero in Figs.~\ref{fig4}) and the FPI mode spectra (marked by the vertical dashed line). 

With further increase of $p_{in}$ and  narrowing of the input field spectra, the maximum of the transmitted FPI spectrum at the input field frequency (at zero in Figs.~\ref{fig4}) increases and narrows as in Figs.~\ref{fig4}c,d. 

If we separate the transmitted field spectra at their local minimum at ${{\omega }_{opt}}-{{\omega }_{l}}=\delta /2$, we see that 48, 70, and 95\% of the transmitted field energy is in the left part, near the maximum of the input field spectrum -- for Fig.~\ref{fig4}b,c, and d correspondingly. So a large amount  of the transmitted field energy is in the  spectral range of the input of a narrow spectrum, as for Figs.\ref{fig4}c and d. The situation is the opposite for a broadband input, as for Figs.\ref{fig4}a, where the principal part of the transmitted field energy is near the maximum of the FPI mode spectrum marked by the vertical dashed line. 

\begin{figure}[thb]
\includegraphics[width=7.0 cm]{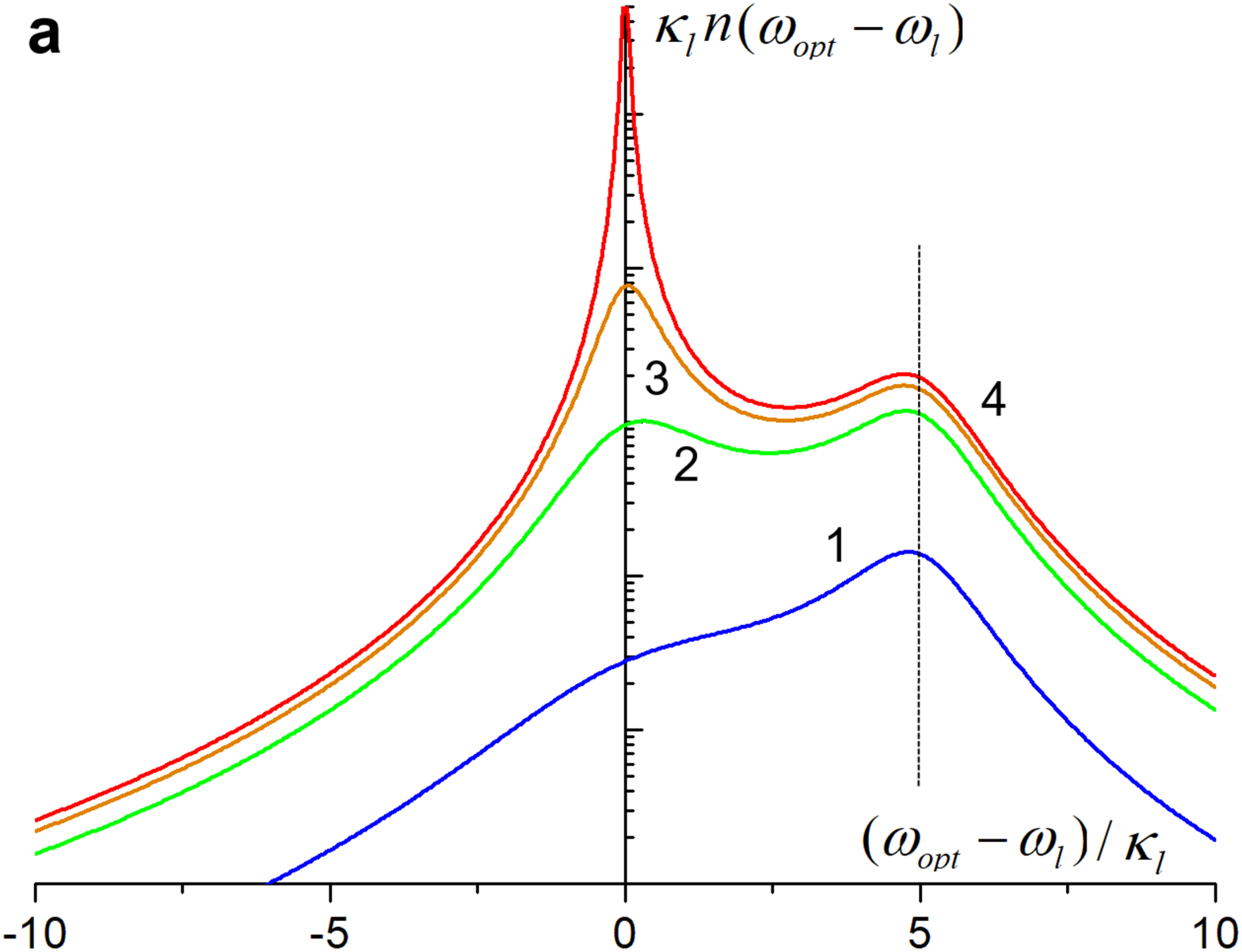}\includegraphics[width=7.0 cm]{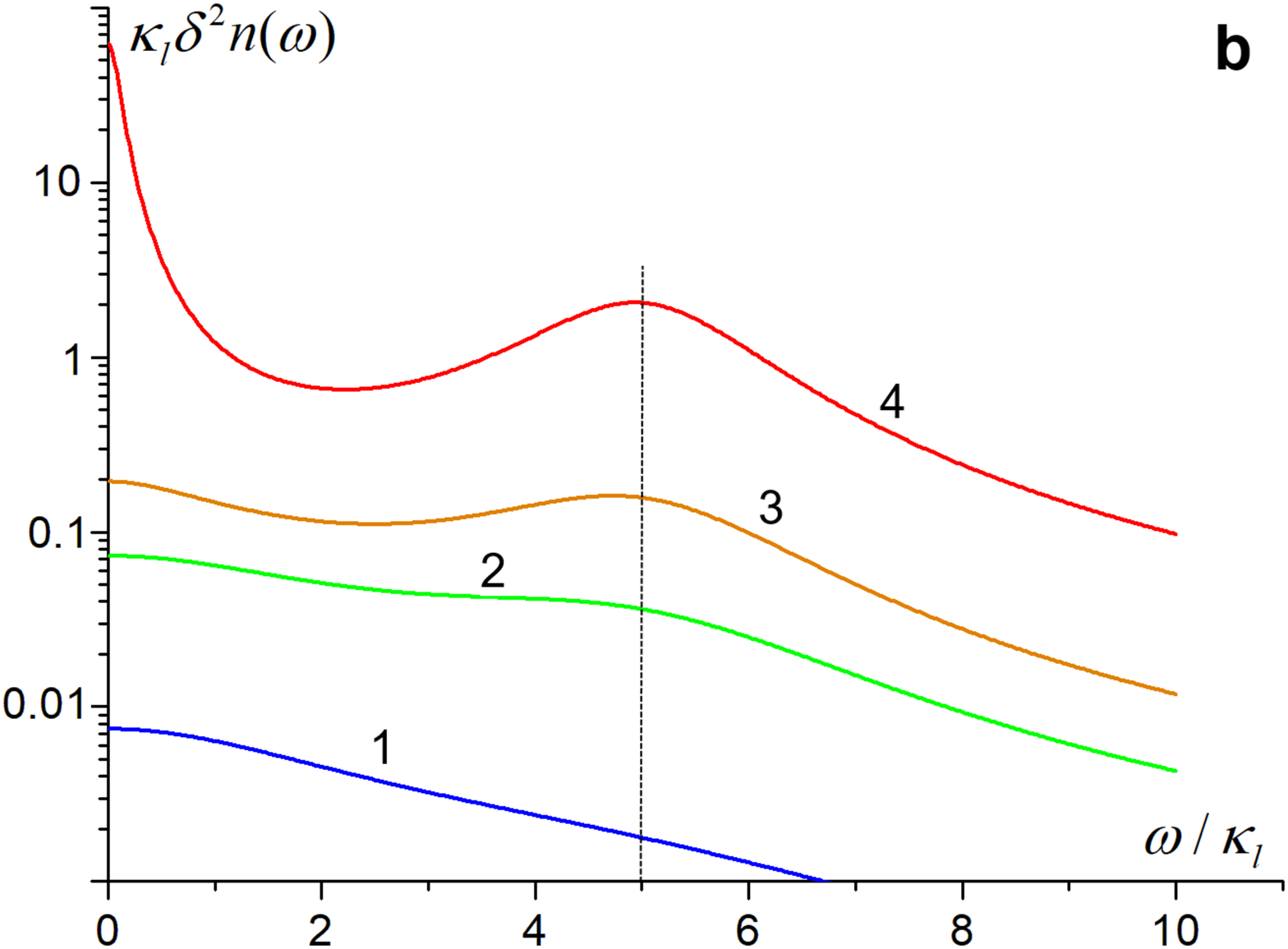}
\caption{(a) The field $n({{\omega }_{opt}}-{{\omega }_{l}})$ and (b) the photon number fluctuation $\delta^2 n(\omega)$ spectra of the FPI cavity mode for ${{p}_{in}}/{{\kappa }_{l}}=0.1$ [${{\gamma }_{l}}/{{\kappa }_{l}}=2.5$] (curves 1); 1.5 [1.1] (2); 5 [0.45] (3); and 50 [0.05] (4). The field spectra have  two maxima, similar to the transmitted field spectra in Fig.~\ref{fig4}. $\delta^2 n(\omega)$ has the sideband maxima at $\omega=\delta$ (shown by the vertical dashed line) in curve 3,4  for a large input field power.}
\label{fig5}
\end{figure}

Fig.~\ref{fig5}a shows the field spectrum $n({{\omega }_{opt}}-{{\omega }_{l}})$ inside the FPI cavity, normalized to $\kappa _{l}^{-1}$ and given by Eq.~\rf{FPI_field_sp}. 
%
%
%
%
\noindent In Fig.~\ref{fig5}, the input field power (the linewidth) increases (decreases) from curve 1 to curve 4. $n({{\omega }_{opt}}-{{\omega }_{l}})$ spectra have  two peaks at large input field, similar to the transmitted field spectrum curves shown  in Figs.~\ref{fig4}. Note that the shape of $n({{\omega }_{opt}}-{{\omega }_{l}})$ curves in Fig.~\ref{fig5}a is different from the well-known Lorenzian (or Airy function) curves of the field mode spectra of the empty FPI \ct{Ismail:16}. 
\subsection{Photon number  fluctuation spectra}
Fig.~\ref{fig5}b shows $\delta^2 n(\omega)$, given by Eq.~\rf{Phot_num_fl_ins}, for the same parameters as for the FPI field spectra in Fig.~\ref{fig5}a. In Fig.~\ref{fig5}b, the width of the input field spectrum exceeds the width of the empty FPI mode for curves 1 and 2, so  $\delta n(\omega)$ is broad with a single maximum at $\omega=0$. When the input power $p_{in}$ increases and the input field spectrum  narrows, the second local peak appears in  $\delta^2 n(\omega)$ at $\omega=\delta$ in curve 3. The peak at $\omega=0$ is higher and narrower while $p_{in}$ increases. The sideband maximum at $\omega=\delta$ grows with $p_{in}$ but not so rapidly as the maximum at $\omega=0$ in curve 4.  

\begin{figure}[thb]
\includegraphics[width=7.0 cm]{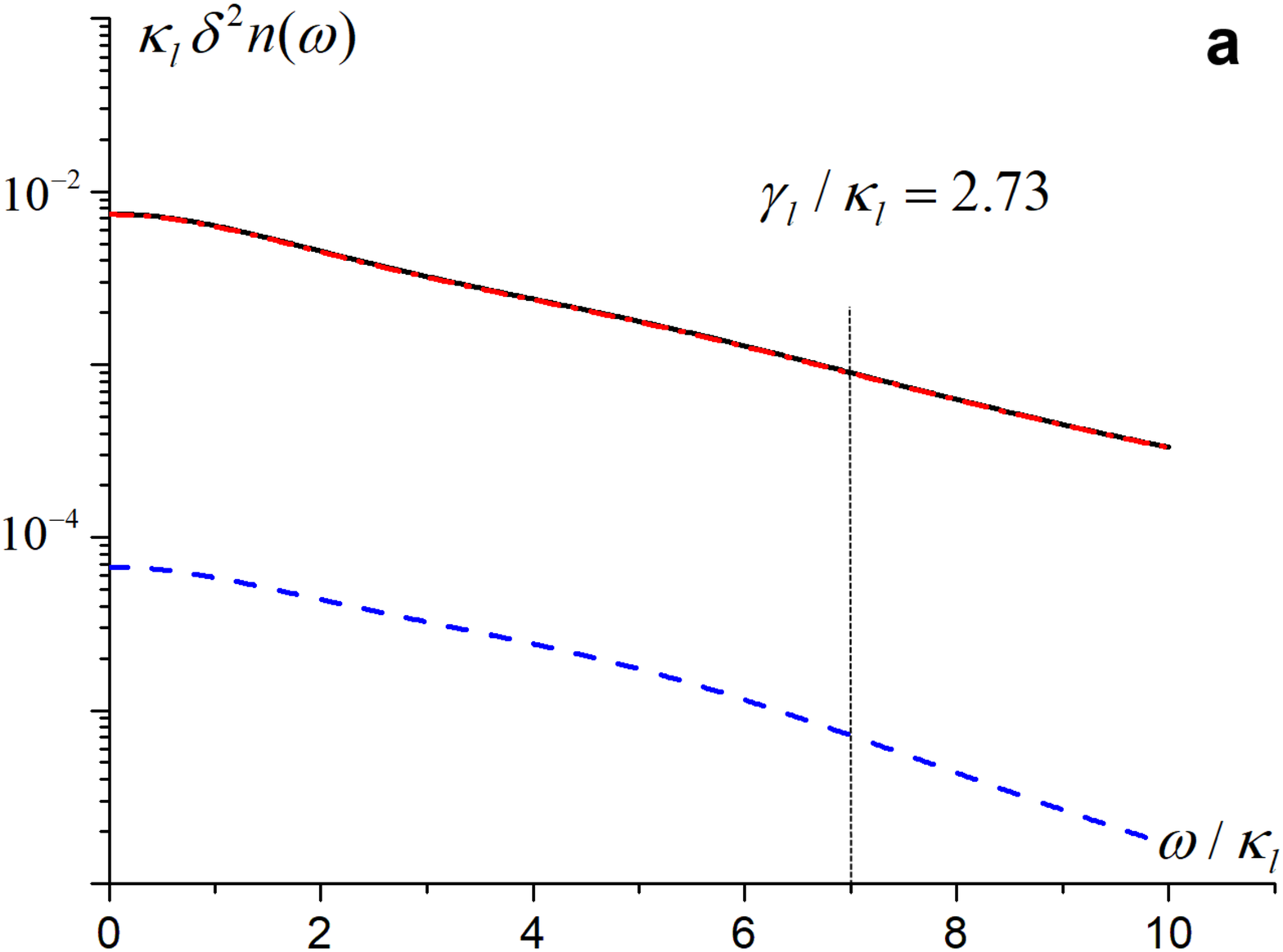}\includegraphics[width=7.0 cm]{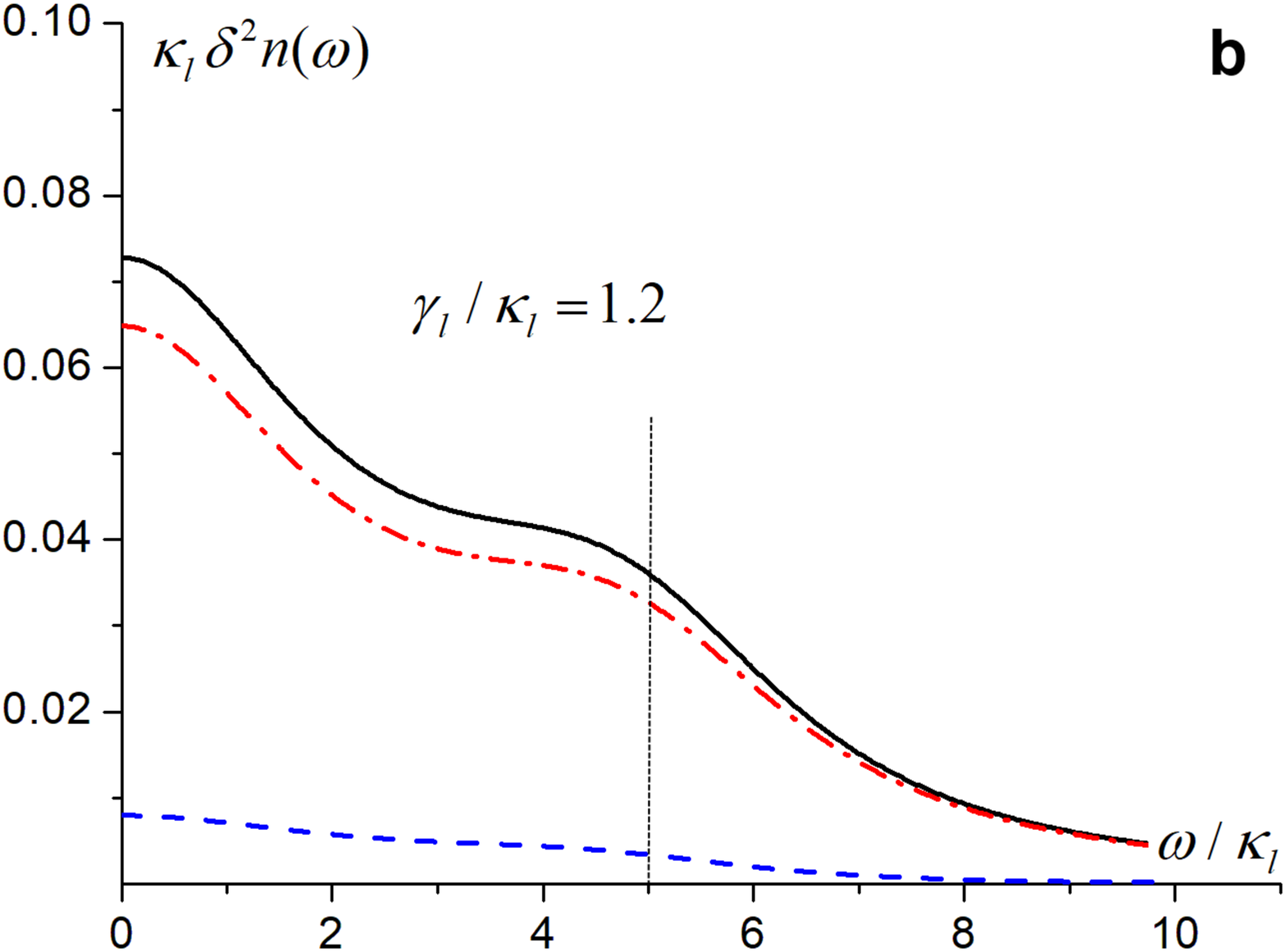}\\
\includegraphics[width=7.0 cm]{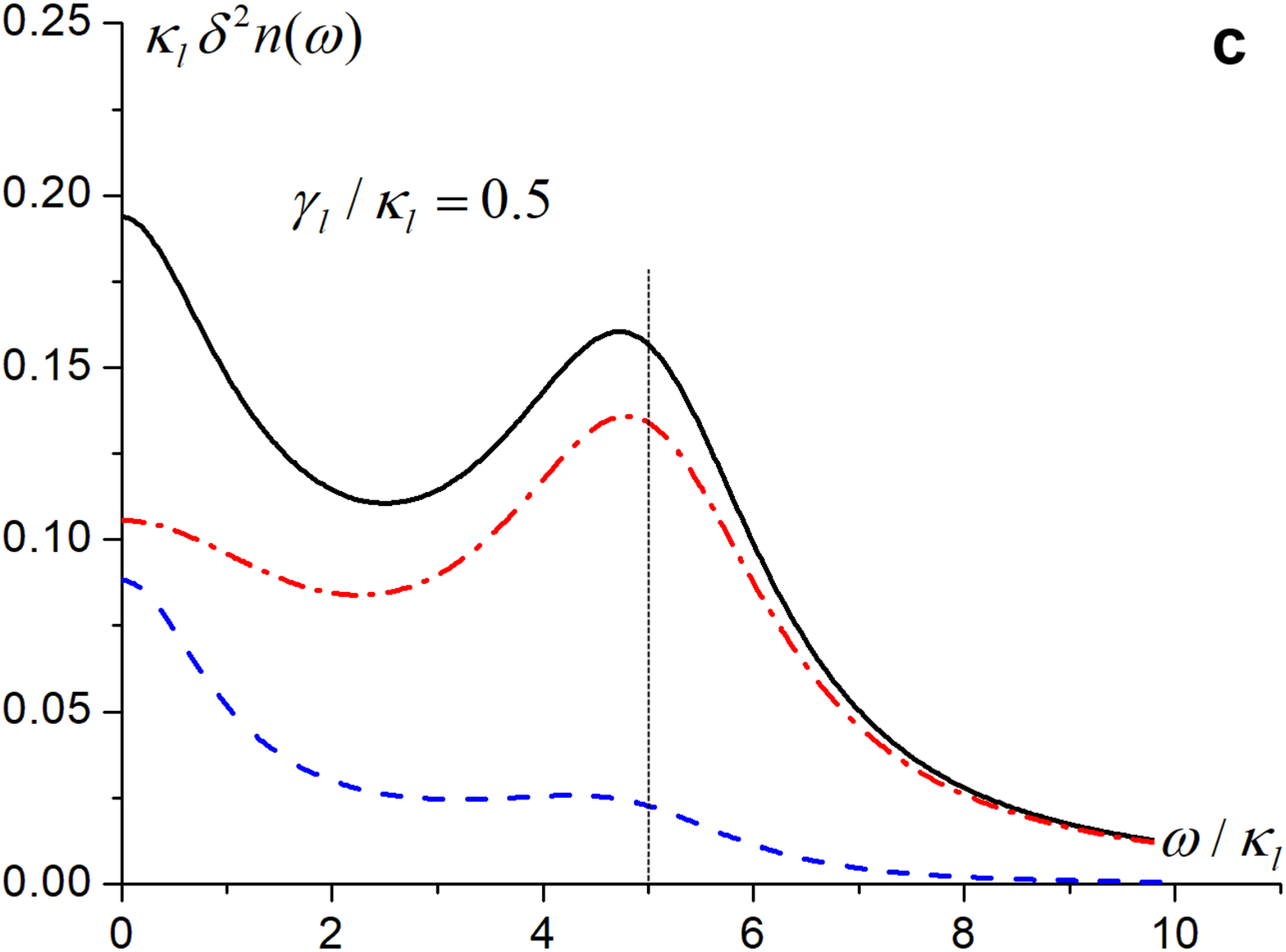}\includegraphics[width=7.0 cm]{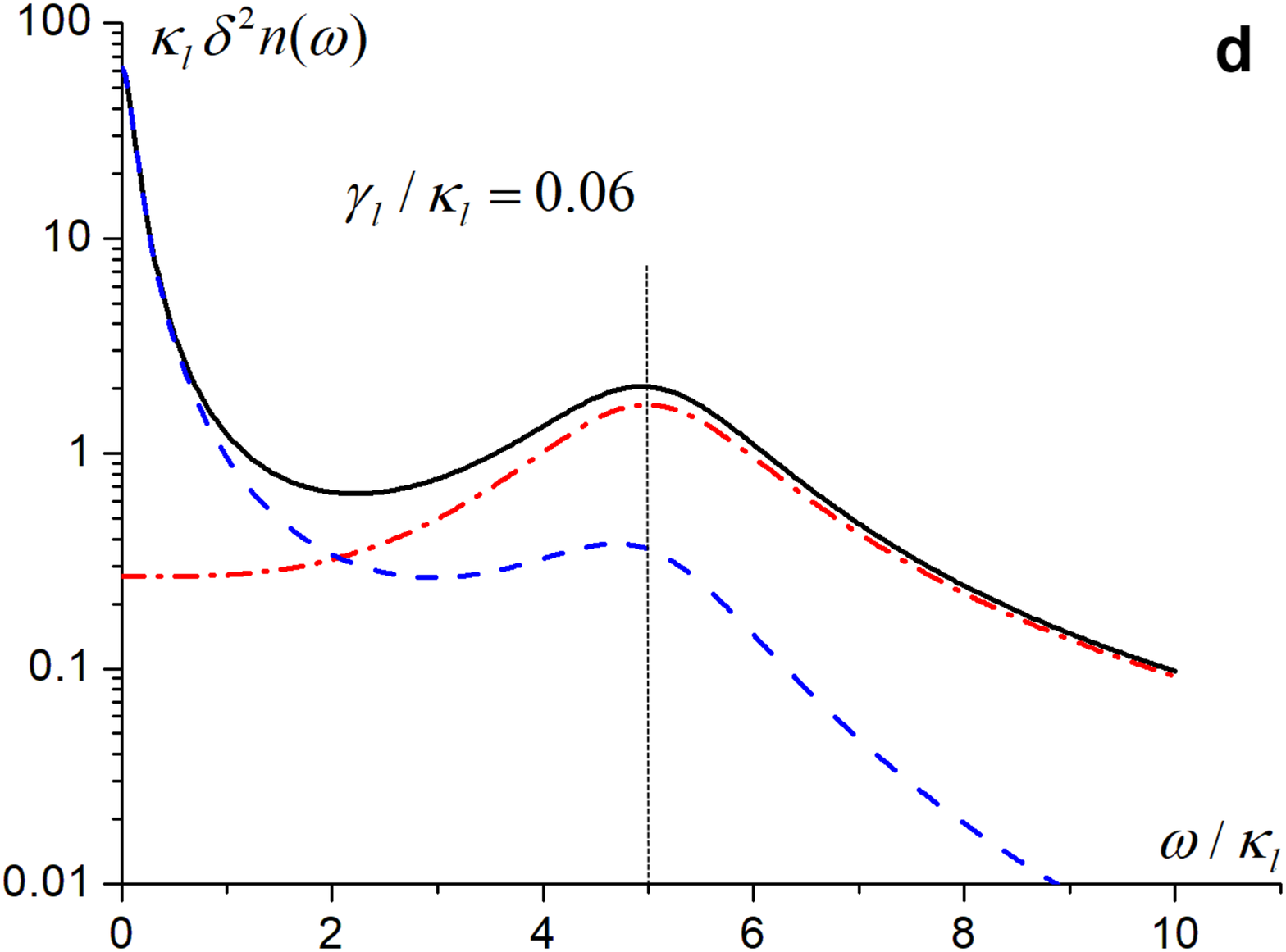}
\caption{Photon number fluctuation spectra in the FPI mode (solid curves) and their classical (dashed curves) and quantum (dash-dotted curves) components. Solid curves  in Figs. a-d are the same as curves 1-4, correspondingly, in Fig.~\ref{fig5}b. The vertical dashed lines mark $\omega=\delta$.}
\label{fig6}
\end{figure}

It is interesting to compare the classical and the quantum contributions in  $\delta^2 n(\omega)$ given by the first and the second terms in Eq.~\rf{Phot_num_fl_ins}, correspondingly. Figs.\ref{fig6}a-d present $\delta^2 n(\omega)$ curves, the same as in Fig.~\ref{fig5}b, together with the quantum and the classical contributions to them. 
%
%
%
%
Fig.~\ref{fig6}a corresponds to a broadband spectrum of the input field with the HWHM $\gamma_l > \kappa_t$. Large quantum fluctuations contribute to $\delta^2 n(\omega)$. The broad $\delta^2 n(\omega)$ spectrum has only one maximum at $\omega=0$.  

With the increase of $p_{in}$ and  the reduction of $\gamma_l$  up to $\gamma_l\sim\kappa_t$, a structure of $\delta^2 n(\omega)$ appears near $\omega=\delta$  in Fig.~\ref{fig6}b; quantum fluctuations still give a major contribution to $\delta^2 n(\omega)$. With further $p_{in}$ growth and the input field spectrum narrowing, the quantum and the classical contributions to $\delta^2 n(\omega)$  become to be  of the same order see the peaks at $\omega=0$ and $\omega=\delta$ in Fig.~\ref{fig6}c. With a large $p_{in}$, when $\gamma_l \ll \kappa_l, \kappa_t$, and the input field source approaches a lasing regime, the spike in $\delta^2 n(0)$ is high and narrow, while the side-band peak in $\omega=\delta$ still presents in Fig.~\ref{fig6}d. Photon number fluctuations in the FPI cavity give a considerable contribution to $\delta^2 n(\omega)$ near the FPI mode maximum at $\omega=\delta$, marked by the vertical dashed  lines in Figs.~\ref{fig6}.

Photon number fluctuations have the maximum at the center of the FPI cavity mode, shifted on $\delta$ respectively to the center of the input field spectrum. That explains the noticeable contribution of quantum fluctuations near  $\omega=\delta$. 

\begin{figure}[thb]
\includegraphics[width=7.0 cm]{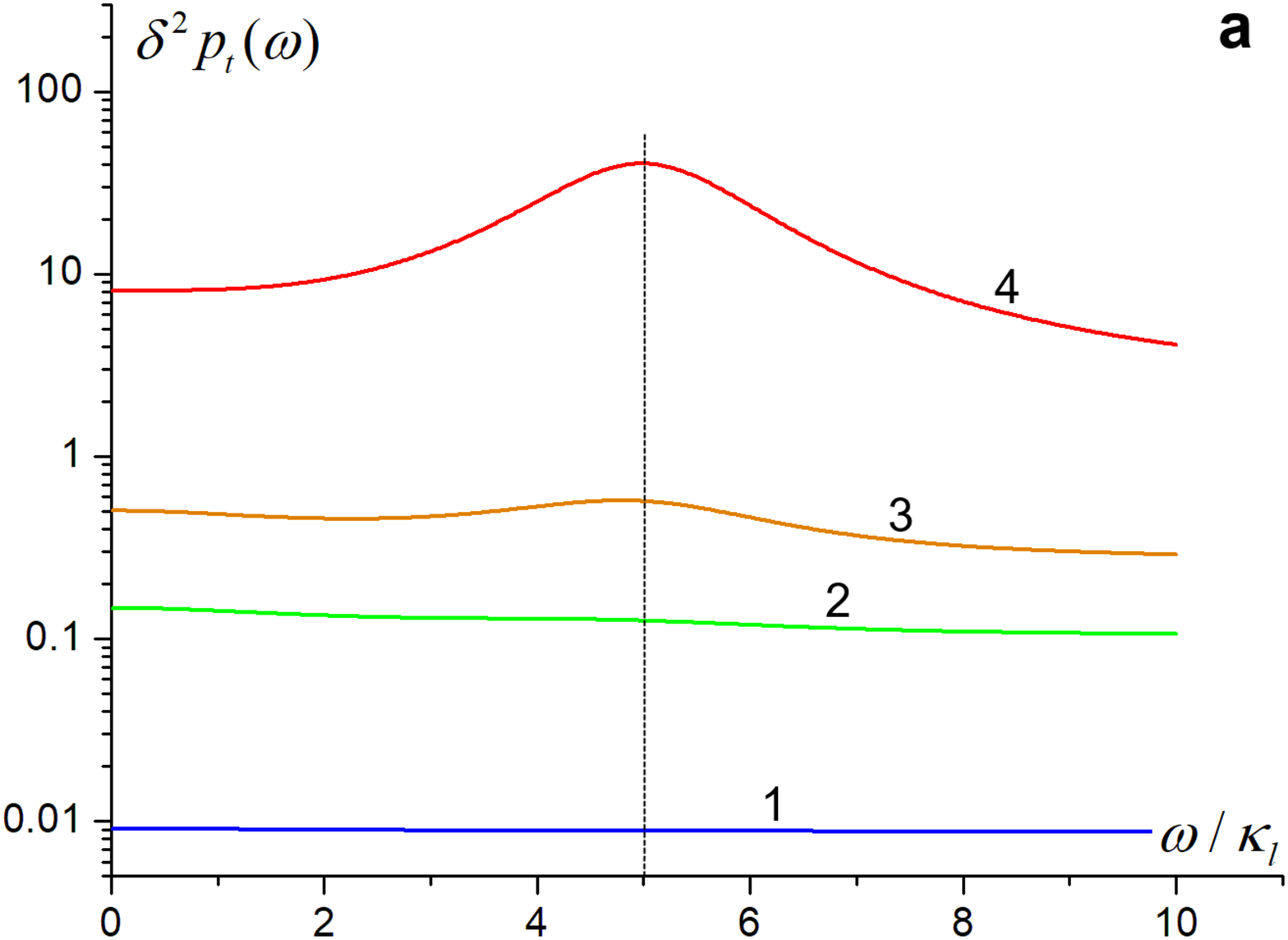}\includegraphics[width=7.0 cm]{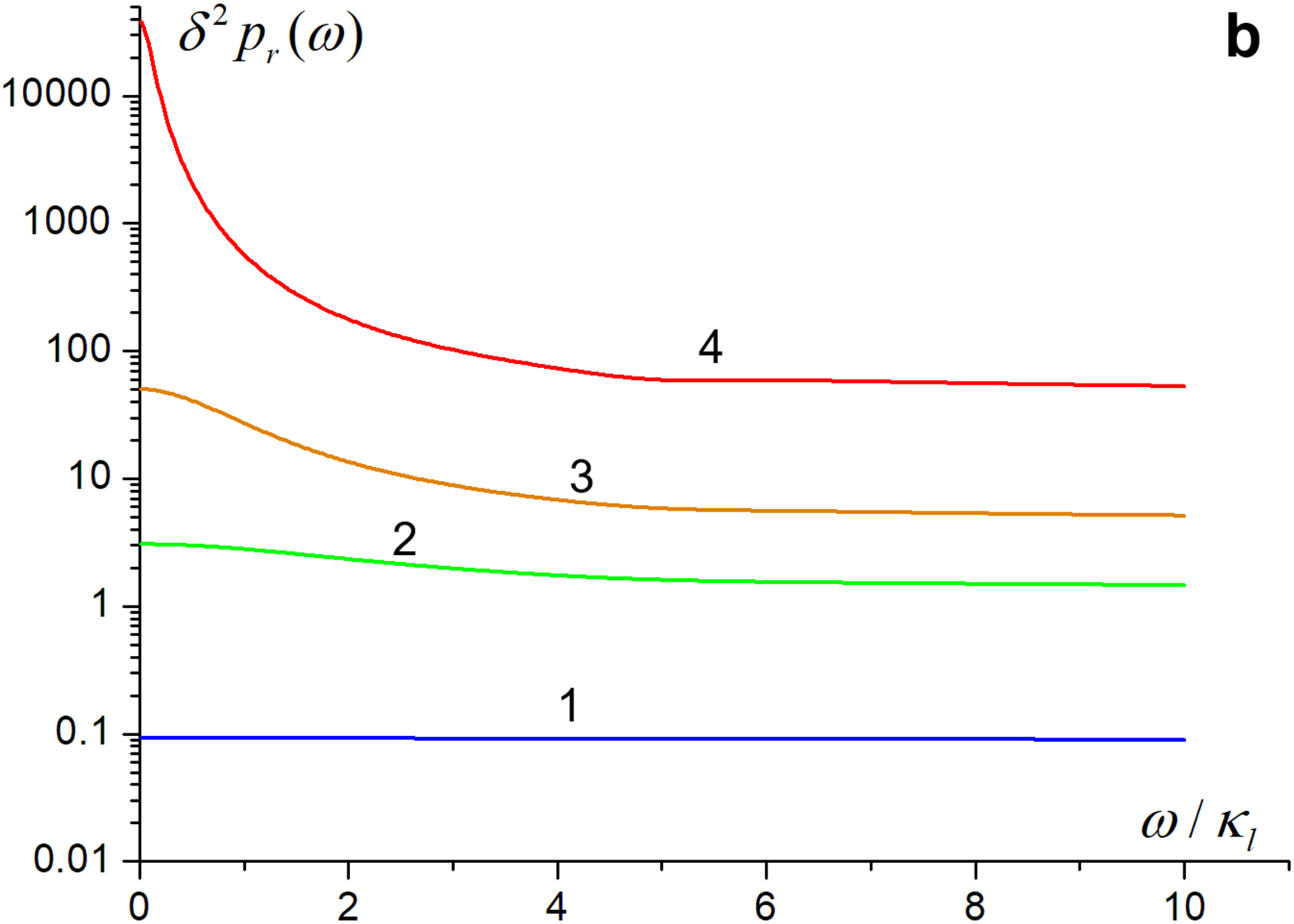}
\caption{Power fluctuation spectra of the transmitted (a) and reflected (b) fields for the same parameters as for curves 1 - 4 in Fig.~\ref{fig5}. The input field power (the HWHM) increases (decreases) from curves 1 to 4. There is only one peak at $\omega=\delta$ in the transmitted field and $\omega=0$ in the reflected field spectra: compare with two peaks in curves inside the FPI shown in Figs.~\ref{fig5}.}
\label{fig7}
\end{figure}

Figs~\ref{fig7} show the power fluctuation spectra of the transmitted  ${{\delta }^{2}}{{p}_{t}}(\omega )$ and the reflected ${{\delta }^{2}}{{p}_{r}}(\omega )$  fields given by Eqs.~\rf{thans_f_sp}, and \rf{RFPF}. Profiles of the spectra different from the power fluctuation spectra inside the FPI in Figs.~\ref{fig5}b and \ref{fig6}.
%
%
%
%
\noindent ${{\delta }^{2}}{{p}_{t}}(\omega )$ has the maximum at $\omega=\delta$, while ${{\delta }^{2}}{{p}_r}(\omega )$ -- at $\omega=0$. Maxima observed if the input field power is large. The quantum part of the transmitted and the reflected power fluctuation spectra does not depend on the frequency, as in Eqs.~\rf{free_sp_final}, \rf{thans_f_sp} and \rf{RFPF}. Quantum contributions only shift the spectra up from the horizontal axis. They do not influence the structure of spectra, which is different with the power fluctuation spectra inside the FPI cavity in Figs.~\ref{fig6}. 
\subsection{Auto-correlation functions}
Inverse Fourier-transforms of  the photon number fluctuation spectrum \rf{Phot_num_fl_ins},  transmitted \rf{thans_f_sp}, and  reflected \rf{RFPF} field power fluctuation spectra lead to auto-correlation functions. Eqs.~\rf{auto_corr_f} and \rf{Phot_num_fl_ins} determine the auto-correlation function ${{\delta }^{2}}n\left( \tau  \right)$ of the FPI cavity photon number fluctuations. The first term in Eq.~\rf{Phot_num_fl_ins} is responsible for the classical component of ${{\delta }^{2}}n\left( \tau  \right)$, and the second is responsible for the quantum component. 

\begin{figure}[thb]
\includegraphics[width=7.0 cm]{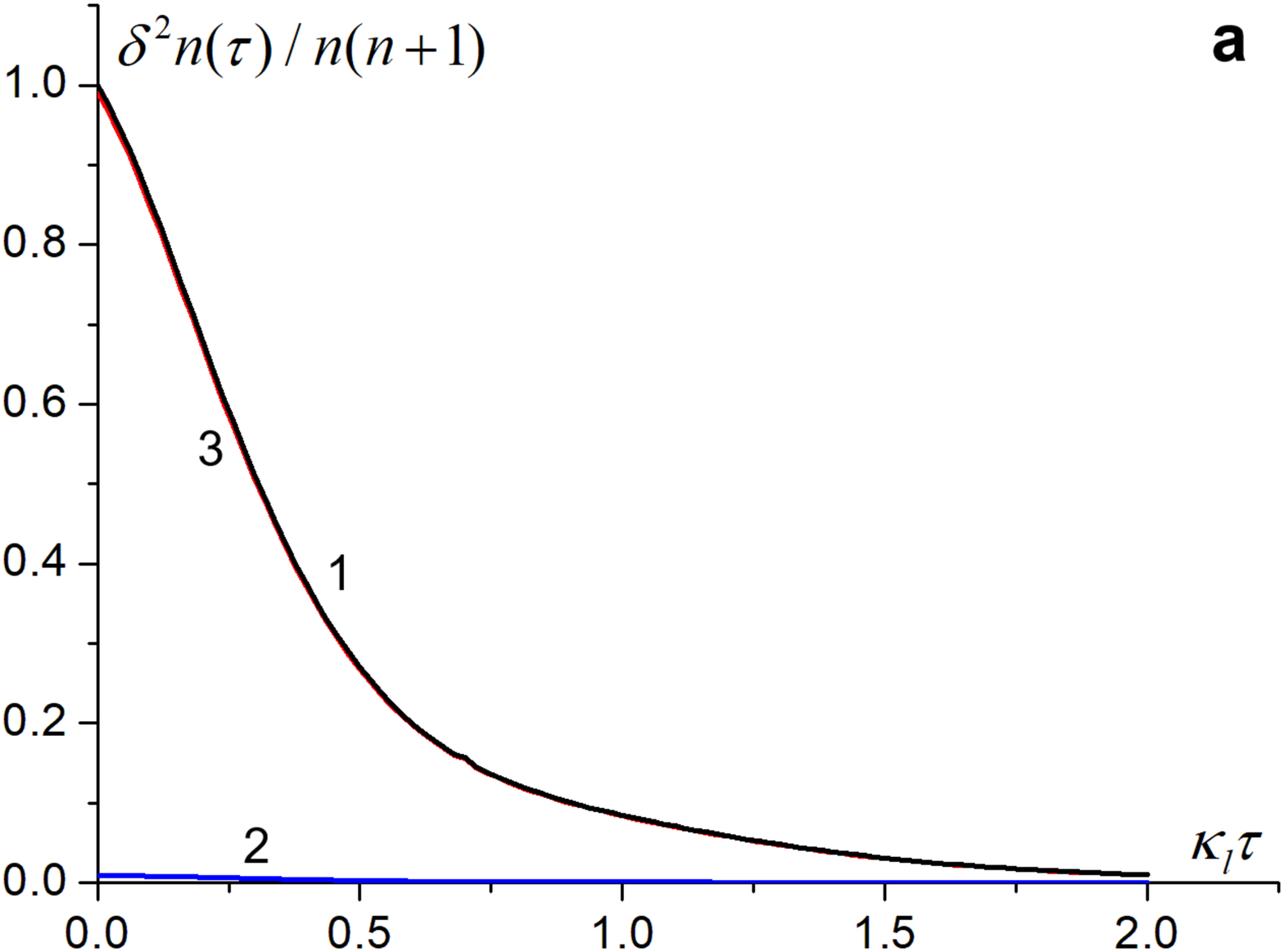}\includegraphics[width=7.0 cm]{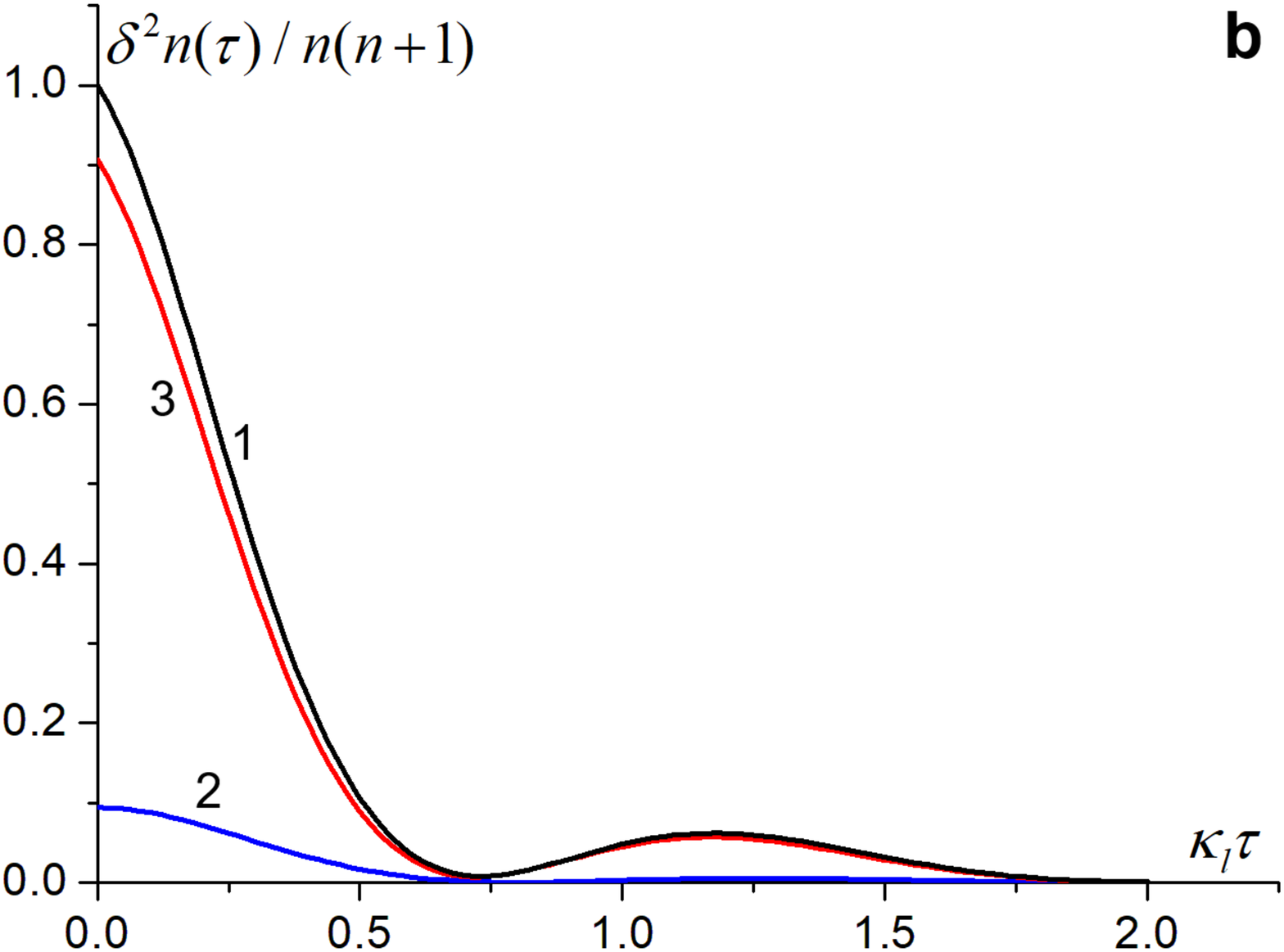}\\\includegraphics[width=7.0 cm]{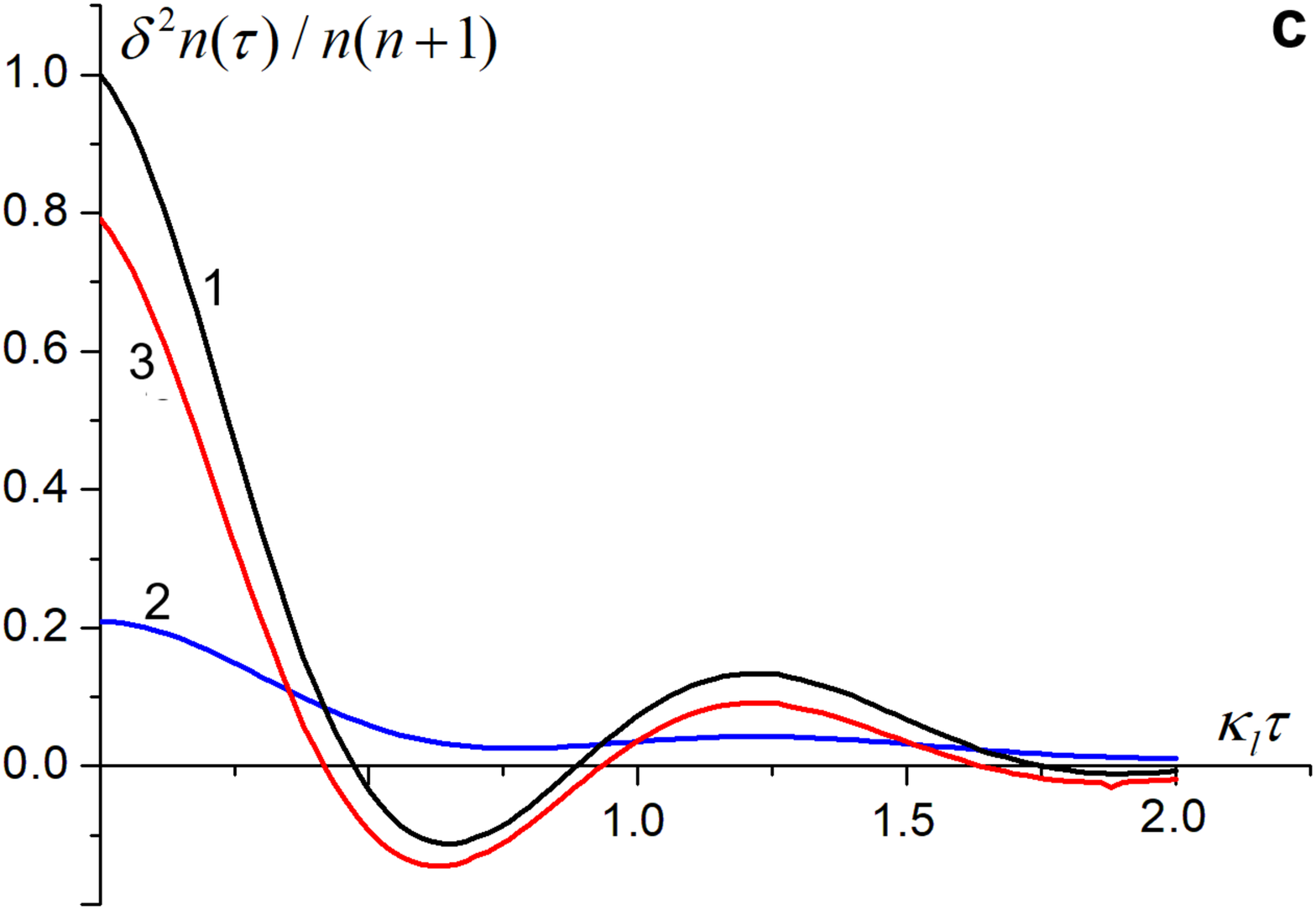}\includegraphics[width=7.0 cm]{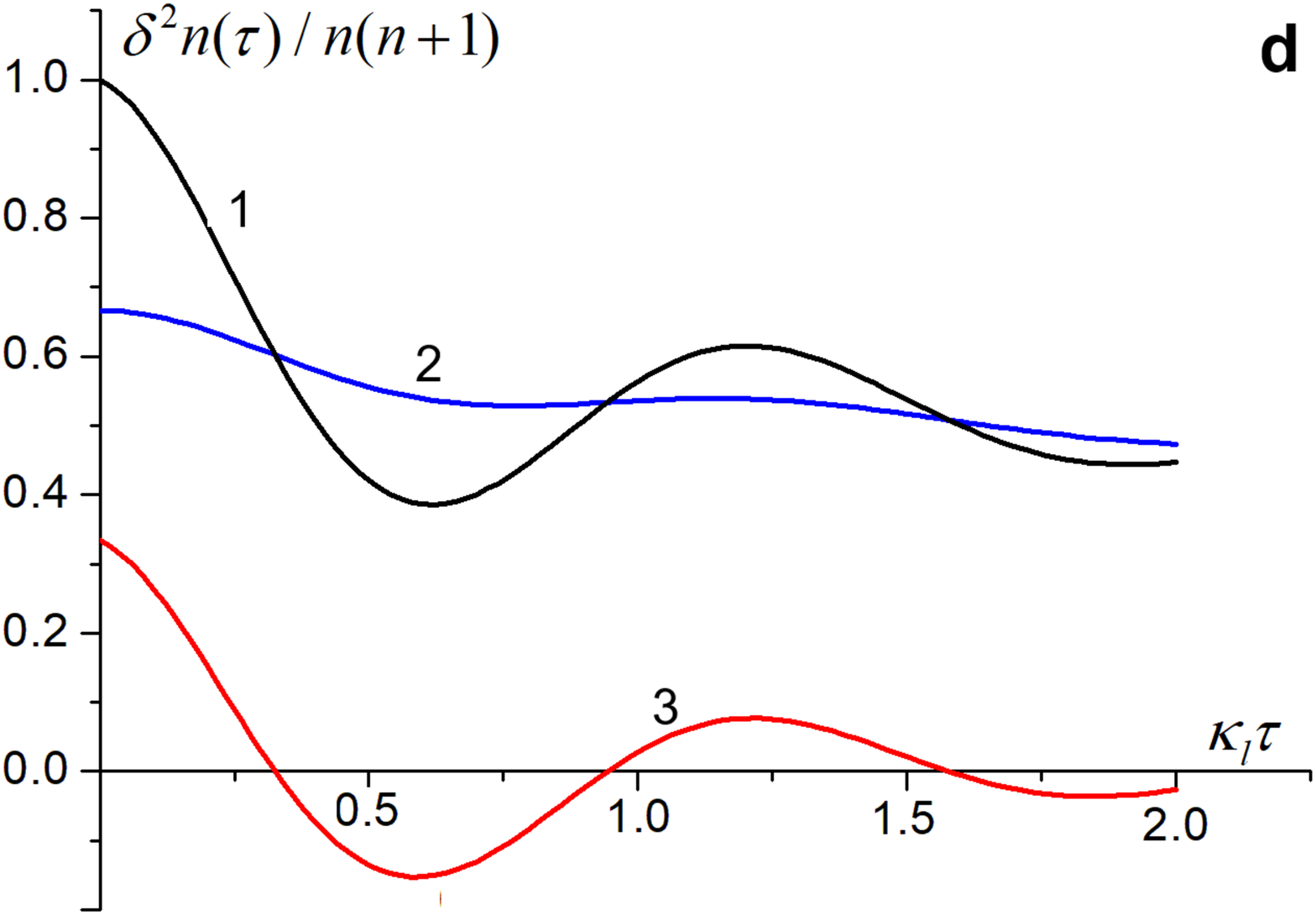}
\caption{Auto-correlation functions for the photon number fluctuations in the FPI cavity (curves 1) consisted of the classical (curves 2) and the quantum (curves 3) contributions. The input field power increases from (a) to (d) with the same values of parameters as for Figs.~\ref{fig6}a-d, correspondingly. }
\label{fig8}
\end{figure}

Fig.~\ref{fig8} shows examples of ${{\delta }^{2}}n\left( \tau  \right)$ and their quantum and classical components. ${{\delta }^{2}}n\left( \tau  \right)$ in Fig.~\ref{fig8}a decreases monotonically with $\tau$ and is determined mostly by quantum fluctuations with a small contribution from classical fluctuations. It is for a small power ${{p}_{in}}/{{\kappa }_{l}}=0.1$ and a wide input field spectrum with HWHM ${{\gamma }_{l}}/{{\kappa }_{l}}=2.73$. ${{\delta }^{2}}n\left( \tau  \right)$ starts to oscillate at larger ${{p}_{in}}/{{\kappa }_{l}}=1.5$ and smaller  the  spectrum HWHM ${{\gamma }_{l}}/{{\kappa }_{l}}=1.2$, as shown  in  Fig.~\ref{fig8}b; ${{\delta }^{2}}n\left( \tau  \right)$ still includes a large part of quantum fluctuations. ${{\delta }^{2}}n\left( \tau  \right)$ oscillates, with its quantum part, and accepts negative values at greater ${{p}_{in}}/{{\kappa }_{l}}=5$ and smaller ${{\gamma }_{l}}/{{\kappa }_{l}}=0.5$ in Fig.~\ref{fig8}c. At large ${{p}_{in}}/{{\kappa }_{l}}=50$ and small ${{\gamma }_{l}}/{{\kappa }_{l}}=0.06$,  ${{\delta }^{2}}n\left( \tau  \right)$ is positive and displays oscillations in the quantum part of ${{\delta }^{2}}n\left( \tau  \right)$, as shown in  Fig.~\ref{fig8}d.

The slow decay of ${{\delta }^{2}}n\left( \tau  \right)$ in Fig.~\ref{fig8}d is related to a narrow line and, correspondingly, long coherency time of the input field.

\begin{figure}[thb]
\includegraphics[width=10.0 cm]{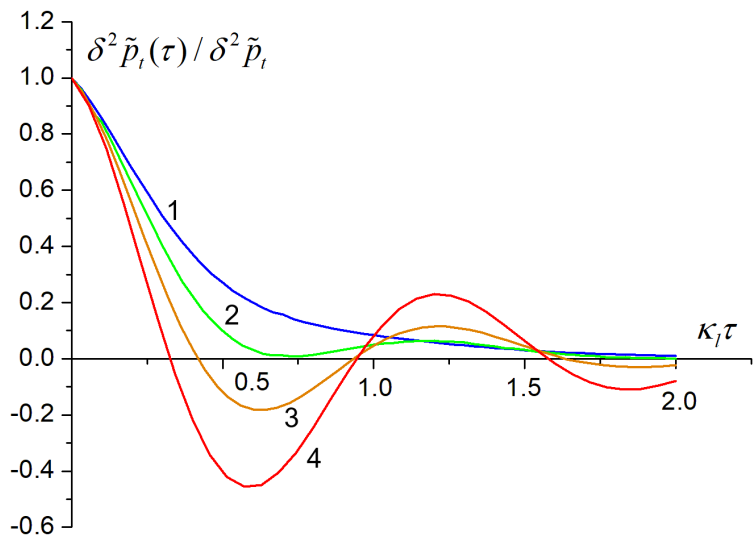}
\caption{Auto-correlation functions for the transmitted field with parameters for curves 1 -- 4 as for Figs.~\ref{fig8}a--d. Delta functions at $\tau=0$, corresponding to quantum noise contributions, are not shown.}
\label{fig9}
\end{figure}

Fig.~\ref{fig9} shows the auto-correlation function ${{\delta }^{2}}{{\tilde{p}}_{t}}(\tau )$ for the transmitted field power fluctuations. ${{\delta }^{2}}{{\tilde{p}}_{t}}(\tau )$ does not contain the term ${{p}_{t}}\delta (\tau)$ presented in  ${{\delta }^{2}}{{{p}}_{t}}(\tau )$ in Eq.~\rf{autocorr_fr_sp_22}. We find ${{\delta }^{2}}{{\tilde{p}}_{t}}(\tau )$  by the inverse Forier-transform of  Eq.~\rf{thans_f_sp} without $2\kappa_2 n$, which corresponds to the quantum part of the transmitted field auto-correlation function. Quantum contributions, proportional to the delta-function $\delta(\tau)$, are not shown in Fig.~\ref{fig9}. Normalizing factor ${{\delta }^{2}}{{\tilde{p}}_{t}}$, used in Fig.~\ref{fig9}, is the integral over frequencies of the expression~\rf{thans_f_sp}, taken without $2\kappa_2 n$.

We see in Fig.~\ref{fig9} that ${{\delta }^{2}}{{\tilde{p}}_{t}}(\tau )$ starts to oscillate and takes negative values with the increase of the input field power $p_{in}$ and the narrowing of the input field spectrum -- from curve 1 to curve 4. Such behavior is related to beatings between the FPI mode and the input field at the non-zero detuning $\delta$ of centers of the FPI mode and the input field spectra.   
%
%
%
%
%
%

The first term in Eq.~\rf{RFPF} dominates in the auto-correlation function ${{\delta }^{2}}{{\tilde{p}}_r(\tau)}$ of the reflected field at chosen parameter values. So ${{\delta }^{2}}{{\tilde{p}}_r(\tau)}$ (taken without delta-function at $\tau=0$) is very well approximated by ${{e}^{-2{{\gamma }_{l}}\left| \tau  \right|}}$, and we do not show ${{\delta }^{2}}{{\tilde{p}}_r}$ on figures. 
\section{Discussion and conclusion}\label{sec4}
We derived formulas \rf{fin_popfl_sp} and \rf{free_sp_final} for the photon number and the transmitted/reflected field power fluctuation spectra. We found transmission/reflection coefficients; transmission, reflection, and the cavity mode field spectra; the photon number (the field power) fluctuation spectra inside (outside) the FPI. We calculate the auto-correlation functions of the FPI -- for the FPI excited by a  finite spectrum width field when maxima of the FPI cavity mode and the input field spectra detuned on $\delta$. We take the detuning $\delta$ several times larger than  HWHMs of spectra of  the FPI mode and the input field. Detuning $\delta$ considered for using the results in the future investigation of the optical bistability \ct{doi:10.1080/00107518308210690, PhysRevA.19.2074} in the miniature FPI with the nonlinear medium  and quantum field with only a few, one, or less than one photon in the cavity. In difference with well-known "macroscopic" FPI \ct{Ismail:16}, quantum fluctuations are significant in the small FPI with a detuning.

We found the transmission $T$ and the reflection $R$ coefficients of the FPI.  $T$ (or $R$) reduced (or increased) with the broadening of the input field spectrum, see  Fig.~\ref{fig3}b.  It is because the finite spectrum width field includes the frequency components shifted from the  resonance with the FPI mode even at the detuning $\delta=0$. The effective FPI mode HWHM  is $\kappa_t+\gamma_l$ in expressions for $T$ and $R$, where $\kappa_t$ and $\gamma_l$ is HWHM of the empty FPI mode  and the input field spectra -- see Eqs.~\rf{R_coef} and \rf{T_coef} for $R$ and $T$.

We investigate the FPI field spectra. The structure of the reflected field spectra (shown by curves 1 in Figs.~\ref{fig4}) remains the same for any width of the input field spectrum. The reflected field spectrum contains the peak at the input field frequency and the gap at the FPI mode frequency. The peak becomes higher and narrower with the narrowing of the input field spectrum. 

The FPI transmitted field spectrum, shown in curves 2 in Figs.~\ref{fig4}, changes its structure with the narrowing of the input field spectrum. The transmitted field has one broad maximum at the FPI mode frequency for broad-band input, see Fig.~\ref{fig4}a. There are two maxima of  similar height, -- when the input field spectral width is of the order of the FPI mode spectral width, see Fig.~\ref{fig4}b,c. The first high and narrow peak is at the input field frequency. The second broad local maximum is at the FPI mode frequency if the input field has high power and a narrow spectrum, see Fig.~\ref{fig4}d.

Fig.~\ref{fig5}a shows the FIP field spectra evolution with the narrowing of the input field spectrum. It is similar to the transmitted field spectra in Fig.~\ref{fig4}. 
It is impossible to separate quantum and classical contributions in the field spectra. 

The two-peak structure in the photon number fluctuation spectra $\delta^2n(\omega)$ inside the FPI appeared with the narrowing of the input field spectrum. We see it in Fig.~\ref{fig5}b. This structure is related to the beating between the input field and the FPI mode field when the input field spectrum is narrower or of the order of the FPI mode spectrum. $\delta^2n(\omega)$ in Fig.~\ref{fig5}b has only the maximum at $\omega=0$ for the weak broad-band input field. 

It is possible to separate the quantum and the classical contributions in the photon number fluctuation spectra inside FPI, such contributions  shown in Figs.~\ref{fig6}.  The broad-band quantum fluctuations dominate in $\delta^2n(\omega)$ with the maximum at $\omega=0$. It is for a weak input field in Fig.~\ref{fig6}a. With the increase of the input field power, the beatings between quantum fluctuations in the FPI mode and the input field appear, leading to the side-band maximum in $\delta^2n(\omega)$ at $\omega=\delta$, see Figs.~\ref{fig6}b,c.  Quantum fluctuations contribute near the FPI mode spectrum maximum. They are high near the local maximum at $\omega=\delta$ in Figs.\ref{fig6}c,d. 

It is not possible to measure $\delta^2n(\omega)$ inside FPI directly. However, the photon number fluctuations influence, for example, the interaction of the quantum field with the nonlinear medium inside the FPI. Photon number fluctuation spectra shown in Figs.~\ref{fig5}, \ref{fig6} help identify the frequency domains where quantum fluctuation dominates. It is meaningful, in particular, if we insert the nonlinear resonant medium inside the small FPI cavity for absorptive bistability  or other purposes; for investigating the quantum field spectra by FPI. 

Field power fluctuation spectra $\delta^2p_t(\omega)$ for the transmitted (Fig.~\ref{fig7}a) and $\delta^2p_r(\omega)$ for the reflected (Fig.~\ref{fig7}b) fields are different from the photon number fluctuation spectra inside FPI in Figs.~\ref{fig5}, \ref{fig6}. $\delta^2p_t(\omega)$ has the maximum at $\omega=\delta$, while $\delta^2p_r(\omega)$ has the maximum at $\omega=0$ if the input field has a high power  and a narrow spectrum.  $\delta^2p_{r,t}(\omega)$ are flat (or almost flat) for a weak and broad-band input field. The white quantum noise does not beat the transmitted (reflected) fields. So there is only one peak in the transmitted (reflected) field power spectra at $\omega=\delta$ ($\omega=0$). 

Auto-correlation functions inside the FPI, shown in Fig.~\ref{fig8}, have the quantum and the classical components. The quantum component dominates at a weak and a broad-band input, and the classical -- at a strong and a narrow spectrum input. We see in Figs.\rf{fig8}b-d oscillations of $\delta^2n(\tau)$ related with a beating between the detuned FPI mode and the input field mode. Beatings disappear at a weak input field and a large quantum noise.  Then $\delta^2n(\tau)$  decays monotonically, as in Fig.~\ref{fig8}a. The auto-correlation function takes negative values  when the contribution of the quantum noise is comparable with the classical part of $\delta^2n(\tau)$, see Fig.~\ref{fig8}c. 

The output field power auto-correlation function $\delta^2p_{t}(\tau)$ in Fig.~\ref{fig9} oscillates, similar to the auto-correlation function inside FPI. $\delta^2p_{t}(\tau)$ accepts negative values when the input field has a high power and a narrow spectrum. Negative values of  $\delta^2p_{t}(\tau)$ are a "trace" of the quantum interference  of the FPI mode noise and the input field  inside the FPI.  The quantum part of $\delta^2p_{t}(\tau)$ is a delta function at $\tau=0$. We do not show this part in Fig.~\ref{fig9}. The difference in quantum contributions to the field power fluctuation spectra inside and outside the FPI related to a broad-band white quantum noise outside and a narrow-band "colored" quantum noise  inside the FPI cavity.  

In conclusion, we carried out quantum analysis of the small, of the size of the wavelength, Fabry-Perot interferometer with a few photons calculating the field, the photon number  (the field power) fluctuation spectra and the second-order time auto-correlation functions.  Results are helpful for the experimental studies of the small FPI used in the photonic integrated circuits, for example, in the delay lines or optical transistors. 
 
\appendix

\section{Quantum input field for the small FPI}\label{Appendix}
The right part of Fig.~\ref{Fig2} shows the scheme of the input field source (a LED or a laser). Operator ${{\hat{a}}_{l}}$ of the field amplitude inside the source cavity satisfies  equations \ct{Andre:19,Protsenko_2021}
\beqr
  {{{\dot{\hat{a}}}}_{l}} &=&-{{\kappa }_{l}}{{{\hat{a}}}_{l}}+\Omega \hat{v}+\sqrt{2{{\kappa }_{l}}}\hat{a}_{in}^{(l)} \lb{MBE}\\
 \dot{\hat{v}}&=&-({{\gamma }_{\bot }}/2)\hat{v}+\Omega f{{{\hat{a}}}_{l}}N+{{{\hat{F}}}_{v}}, \nonumber
\eeqr
where $\hat{v}$ is the active medium polarization operator, $\Omega$ is the vacuum Rabi frequency, ${{\gamma }_{\bot }}/2$ is the polarization decay rate, factor $f=1/2$, the population inversion $N=2{{N}_{e}}-{{N}_{0}}$, ${{N}_{e}}$ (${{N}_{g}}$) is the population of the upper (the low) states of the active medium with $N_0$ two-level emitters, $\hat{F}_{\nu}$ is the Langevin force with non-zero correlations
\beq
\left\langle {{{\hat{F}}}_{{{v}^{+}}}}(-\omega ){{{\hat{F}}}_{v}}(\omega ') \right\rangle ={{\gamma }_{\bot }}f{{N}_{e}}\delta (\omega +\omega '), \hspace{0.5cm}  \left\langle {{{\hat{F}}}_{v}}(\omega '){{{\hat{F}}}_{{{v}^{+}}}}(-\omega ) \right\rangle ={{\gamma }_{\bot }}f{{N}_{g}}\delta (\omega +\omega '). \lb{corr_app}
\eeq
We suppose that the input field is not too large and, following \ct{Andre:19, Protsenko_2021}, neglect in Eqs.~\rf{MBE} fluctuations of populations of states of the two-level medium. 

We take the source with ${{\kappa }_{l}}\ll {{\gamma }_{\bot }}/2$ and eliminate the polarization $\hat{v}$ adiabatically, setting $\dot{\hat{v}}=0$ in Eqs.~\rf{MBE}. We obtain $\Omega \hat{v}=\left( {{\kappa }_{l}}N/{{N}_{th}} \right){{\hat{a}}_{l}}+\left( 2\Omega /{{\gamma }_{\bot }} \right){{\hat{F}}_{v}}$ and
\beq
{{\dot{\hat{a}}}_{l}}=-{{\kappa }_{l}}\eta {{\hat{a}}_{l}}+\left( 2\Omega /{{\gamma }_{\bot }} \right){{\hat{F}}_{v}}+\sqrt{2{{\kappa }_{l}}}\hat{a}_{in}^{(l)},\lb{Eq_field_s}
\eeq
where $\eta =1-N/{{N}_{th}}$, $N_{th}=\kappa_l\gamma_{\perp}/2\Omega^2f$ is the threshold population inversion found in the semi-classical laser theory \ct{trove.nla.gov.au/work/21304573,Andre:19}. Eq.~\rf{Eq_field_s} leads to the Fourier-component operator of the field inside the source cavity
\beq
{{\hat{a}}_{l}}(\omega )=\frac{\left( 2\Omega /{{\gamma }_{\bot }} \right){{{\hat{F}}}_{v}}(\omega )+\sqrt{2{{\kappa }_{l}}}\hat{a}_{in}^{(l)}(\omega )}{{{\kappa }_{l}}\eta -i\omega }\lb{FC_source_cav}
\eeq
Operators $\hat{a}_{in}^{(l)}$ and $\hat{a}_{out}^{(l)}$ of amplitudes of the input vacuum and the output fields at the output (the left) mirror of the source in Fig.~\ref{Fig2} satisfies the boundary condition $\hat{a}_{out}^{(l)}+\hat{a}_{in}^{(l)}=\sqrt{2{{\kappa }_{l}}}{{\hat{a}}_{l}}$. Taking this condition and Eq.~\rf{FC_source_cav}, we find the Fourier component of the FPI input field
\beq
{{\hat{a}}_{in}}(\omega )=\hat{a}_{out}^{(l)}(\omega )=\frac{\sqrt{2{{\kappa }_{l}}}\left( 2\Omega /{{\gamma }_{\bot }} \right){{{\hat{F}}}_{v}}(\omega )+\left[ {{\kappa }_{l}}(1+N/{{N}_{th}})+i\omega  \right]\hat{a}_{in}^{(l)}(\omega )}{{{\kappa }_{l}}\eta -i\omega } \lb{input_FPI_f_fc}
\eeq
Using the relation $\left\langle \hat{a}_{out}^{(l)+}(-\omega )\hat{a}_{out}^{(l)}(\omega ') \right\rangle ={{p}_{in}}(\omega )\delta (\omega +\omega ')$,  Eq.~\rf{input_FPI_f_fc} and the first correlation in Eq.~\rf{corr_app} we find the FPI input field power Fourier-component
\beq
{{p}_{in}}(\omega )=\frac{4\kappa _{l}^{2}{{N}_{e}}/{{N}_{th}}}{{{\left( {{\kappa }_{l}}\eta  \right)}^{2}}+{{\omega }^{2}}}\lb{FPI_inp_0}
\eeq
The FPI input field power is ${{p}_{in}}=({2\pi })^{-1}\int\limits_{-\infty }^{\infty }{{{p}_{in}}(\omega )d\omega }$. We represent ${{p}_{in}}(\omega )$ as a function of $p_{in}$
\beq
{{p}_{in}}(\omega )={{p}_{in}}L(\omega ,{{\gamma }_{l}}),\lb{sp_short}
\eeq
 and use $p_{in}$ as a governing parameter in calculations of spectra in the main text. In Eq.~\rf{sp_short} 
\beq
{{\gamma }_{l}}=\frac{{{\gamma }_{\max }}}{1+{{p}_{in}}/{{\kappa }_{l}}} \lb{gamma_las}
\eeq
is the HWHM of the input field spectrum. The maximum value of ${{\gamma }_{l}}$ is ${{\gamma }_{\max }}={{\kappa }_{l}}\left( 1+{{N}_{0}}/{{N}_{th}} \right)$; ${{\gamma }_{l}}\to {\gamma }_{\max }$ at ${{p}_{in}}\to 0$.



\bibliography{myrefs}

\end{document}